\documentclass[12pt]{article}  
  
\usepackage{epsfig,epsf}  
\usepackage{graphics}  
\usepackage{cite}  
\usepackage{geometry}
\usepackage{xcolor} 

\usepackage{amsmath}  
\usepackage{amsthm}  
\usepackage{amsfonts}  
\usepackage{amssymb} 
\usepackage{bm} 

\usepackage{slashed}
\usepackage{braket}

\usepackage{hyperref}
\usepackage{booktabs}

\newcommand{\p}{\partial}
\newcommand{\MS}{\ensuremath{\overline{\text{MS}}}}
\newcommand{\LQCD}{\Lambda_\text{QCD}}
\DeclareMathOperator{\im}{Im}

\newcounter{MBQ}

\begin{document}
\allowdisplaybreaks
\thispagestyle{empty}  
  
\begin{flushright}  
{\small  
IPPP/18/20 \\[0.1cm]
\today
}  
\end{flushright}  
  
\vskip2cm  
\begin{center}  
\textbf{\Large\boldmath Higher-order condensate corrections to $\Upsilon$ 
masses, leptonic decay rates and sum rules}
\\  
\vspace{2cm}  
{\sc T.~Rauh}\\[0.5cm]  
\vspace*{0.5cm} 
% {\it 
% Physik Department T31,\\ 
% James-Franck-Stra\ss{}e~1,
% Technische Universit\"at M\"unchen,\\
% 85748 Garching, Germany}\\[0.5cm] 
{\it 
IPPP, Department of Physics,
University of Durham,\\
DH1 3LE, United Kingdom}
  
\def\thefootnote{\arabic{footnote}}  
\setcounter{footnote}{0}  
  
\vskip3cm  
\textbf{Abstract}\\  
\vspace{1\baselineskip}  
\parbox{0.9\textwidth}{
With the recent completion of NNNLO results, the perturbative description 
of the $\Upsilon$ system has reached a very high level of sophistication. 
We consider the non-perturbative corrections as an expansion in terms of 
local condensates, following the approach pioneered by Voloshin and Leutwyler. 
The leading order corrections up to dimension eight and the potential NLO 
corrections at dimension four are computed and given in analytical form. 
We then study the convergence of the expansion for the masses, the leptonic 
decay rates and the non-relativistic moments of the $\Upsilon$ system. 
We demonstrate that the condensate corrections to the $\Upsilon(1S)$ mass 
exhibit a region with good convergence, which allows us to extract 
$\overline{m}_b(\overline{m}_b) = 4214\pm37\,(\text{pert.})\,_{-22}^{+20}\,(\text{non-pert.})\text{ MeV}$, 
and show that non-perturbative contributions to the moments with 
$n\approx10$ are negligible.}  
  
\end{center}  
  
%{\small \tableofcontents}  
  
\newpage  
\setcounter{page}{1}

%%%%%%%%%%%%%%%%%%%%%%%%%%%%%%%%%%%%%%%%%%%%%%%%%%%%%%%%%%%%%%%%%%%%%%%%%%%%%%%%%%%%%%%%%%
%%%%%%%%%%%%%%%%%%%%%%%%%%%%%%%%     Introduction      %%%%%%%%%%%%%%%%%%%%%%%%%%%%%%%%%%%
%%%%%%%%%%%%%%%%%%%%%%%%%%%%%%%%%%%%%%%%%%%%%%%%%%%%%%%%%%%%%%%%%%%%%%%%%%%%%%%%%%%%%%%%%%

\section{Introduction}

In recent years, the accuracy of the perturbative description of the bottomonium system 
has been extended to next-to-next-to-next-to-leading order (NNNLO). The full 
spectrum~\cite{Kiyo:2013aea,Kiyo:2014uca}\footnote{See also \cite{Peset:2015vvi} for the 
case of unequal masses.}, the leptonic decay rate of the $\Upsilon(1S)$~\cite{Beneke:2014qea} 
and the non-relativistic moments of the total $e^+e^-\to b\bar{b}X$ cross 
section~\cite{Beneke:2014pta} have been determined and perturbation theory is well behaved. 
This has important phenomenological implications. For instance, some of the most precise 
determinations of the bottom-quark mass rely on the comparison of the perturbative expressions 
for the non-relativistic moments~\cite{Beneke:2014pta,Beneke:2016oox,Penin:2014zaa,Hoang:2012us} 
or the masses of the $\Upsilon(1S)$ and $\eta_b(1S)$ resonances~\cite{Ayala:2014yxa,Ayala:2016sdn,Kiyo:2015ufa} 
and recently also the $n=2$ states~\cite{Mateu:2017hlz} with their experimental values. 

Bottomonium can be treated as a non-relativistic system where the bottom-quark velocity $v$ 
is of the order of the strong-coupling constant: $v\sim\alpha_s(m_b v)\ll1$. There is a 
large hierarchy between the dynamical scales $m_b$ (hard), $m_b v$ (soft) and $m_bv^2$ 
(ultrasoft) of the system. The perturbative calculations~\cite{Kiyo:2013aea,Kiyo:2014uca,Peset:2015vvi,Beneke:2014qea,Beneke:2014pta} 
were performed using the effective theory \emph{potential non-relativistic QCD} 
(PNRQCD)~\cite{Pineda:1997bj,Beneke:1999qg,Brambilla:1999xf,Beneke:2013jia}, where the hard 
and the soft scale have been integrated out and the only dynamical modes left are potential 
bottom quarks $\psi$ and anti-bottom quarks $\chi$, with energy and momentum of the order 
$m_b v^2$ and $m_b v$, respectively, as well as ultrasoft gluons and light quarks. 
The Lagrangian for perturbative calculations up to NNNLO takes the form 
\begin{equation}
 \begin{aligned}
  \mathcal{L}_{\text{PNRQCD}}=&\hspace*{0.15cm}\psi^\dagger\left(i\p_0+\frac{\boldsymbol{\p}^2}{2m}+g_sA_0(t,\mathbf{0})-g_s\mathbf{x}\cdot\mathbf{E}(t,\mathbf{0})+\frac{\boldsymbol{\p}^4}{8m^3}\right)\psi\\
  &+\chi^\dagger\left(i\p_0-\frac{\boldsymbol{\p}^2}{2m}+g_sA_0(t,\mathbf{0})-g_s\mathbf{x}\cdot\mathbf{E}(t,\mathbf{0})-\frac{\boldsymbol{\p}^4}{8m^3}\right)\chi\\
  &+\int d^{d-1}\mathbf{r}\left[\psi_a^\dagger\psi_b\right](x+\mathbf{r})V_{ab;cd}(\mathbf{r},\p)\left[\chi_c^\dagger\chi_d\right](x)\\
  &+\mathcal{L}_\text{ultrasoft}, 
 \end{aligned}
\label{eq:LPNRQCD}
\end{equation}
where the coupling to the ultrasoft gluon field in the bottom-quark bilinear parts has been 
multipole expanded in the spatial components~\cite{Beneke:1999zr}, the third line describes 
the interactions through spatially non-local potentials, which are given 
in~\cite{Beneke:2013jia,Lee:2016cgz}, and the ultrasoft Lagrangian is a copy of the QCD 
Lagrangian which only contains the ultrasoft gluon and light quark fields. 

Purely perturbative calculations within PNRQCD are valid when the ultrasoft scale $m_b v^2$ 
is much larger than the QCD scale $\LQCD$. This is certainly the case for 
top quarks, which are studied in~\cite{Beneke:2015kwa,Beneke:2017rdn}\footnote{Furthermore, 
the sizeable top-quark decay width provides a cutoff on non-perturbative effects~\cite{Bigi:1986jk,Fadin:1987wz}.}, 
but is questionable in the bottomonium sector. 
Assuming the hierarchy holds, non-perturbative corrections can be incorporated in terms of 
local vacuum condensates as a power series in $(\LQCD/(m_b v^2))^2$, following the approach of 
Voloshin and Leutwyler~\cite{Voloshin:1978hc,Voloshin:1979uv,Leutwyler:1980tn,Voloshin:1995sf}. 
In this work, we compute higher-order corrections in this approach and assess the convergence 
of the series. 

In the limit $\LQCD\ll m_b v^2$, the gluon field in the PNRQCD Lagrangian 
can be split into two parts 
\begin{equation}
 A_\mu(t,\mathbf{x})=A_\mu^\text{us}(t,\mathbf{x}) + A_\mu^\text{np}(t,\mathbf{x}).
\end{equation}
The superscripts denote the ultrasoft and the non-perturbative gluon field with momentum 
of the order $m_b v^2$ and $\LQCD$, respectively. All couplings of the non-perturbative 
component to other modes must be multipole-expanded because the non-perturbative field 
with a large wavelength of the order $1/\LQCD$ cannot resolve the dynamics of the potential 
bottom quarks or of the ultrasoft gluons. 
A convenient gauge choice for the non-perturbative gluon field is given by Fock-Schwinger 
gauge 
\begin{equation}
  \mathbf{x}\cdot\mathbf{A}^\text{np}(t,\mathbf{x})=0,\hspace{1cm}A_0^\text{np}(t,\mathbf{0})=0,
  \label{eq:FockSchwingerGauge}
\end{equation}
which removes the coupling of the bottom quarks to the $A_0^\text{np}$ field. The leading 
non-perturbative contribution in the PNRQCD Lagrangian then takes the form of a 
chromoelectric dipole term 
\begin{equation}
 \mathcal{L}_\text{non-perturbative}=\psi^\dagger\left(-g_s\mathbf{x}\cdot\mathbf{E}^\text{np}(0,\mathbf{0})+\dots\right)\psi 
 +\chi^\dagger\left(-g_s\mathbf{x}\cdot\mathbf{E}^\text{np}(0,\mathbf{0})+\dots\right)\chi,
 \label{eq:Lnonpert}
\end{equation}
and is of the order $m_b^2v^2\LQCD^2$ because $\mathbf{E}^\text{np}\sim\LQCD^2$ and the 
strong coupling at the QCD scale is counted as order one. This implies that the 
chromoelectric dipole coupling to the non-perturbative gluon field is suppressed by 
$v(\LQCD/(m_bv^2))^2$ with respect to the leading order Lagrangian. 
The time-dependent terms in the multipole expansion and the expanded couplings between 
the non-perturbative and ultrasoft modes in $\mathcal{L}_\text{ultrasoft}$ are not 
required for the leading order condensate corrections. Their relevance at higher orders 
is assessed in Section~\ref{sec:dim4NLO}, where we discuss the NLO QCD corrections to 
the leading term in the Voloshin-Leutwyler approach. 

The condensate corrections to the considered observables can be extracted from the 
non-relativistic Green function at the origin 
\begin{equation}
 G(E) \equiv \hm{\langle}\hspace{-0.2cm}\hm{\langle}\mathbf{0}\hm{|}\hspace{-0.12cm}\hm{|}\left(\hat{H}-E-i0\right)^{-1}\hm{|}\hspace{-0.12cm}\hm{|}\mathbf{0}\hm{\rangle}\hspace{-0.2cm}\hm{\rangle}, 
 \label{eq:GE_def}
\end{equation}
where $E=\sqrt{s}-2m_b$ is the non-relativistic energy of the system and the Hamiltonian has the form 
\begin{equation}
 \hat{H} = \hat{H}_{b\bar{b}} + \hat{H}_\text{np} + \hat{H}_D +\dots\;.
\end{equation}
The bottomonium Hamiltonian follows from~\eqref{eq:LPNRQCD} and is given by a perturbative 
series in $g_s\sim v^{1/2}$: 
\begin{equation}
 \hat{H}_{b\bar{b}} = \sum\limits_{i=0,\frac12,1,\dots} \hat{H}_{b\bar{b},i} = -\frac{\nabla^2}{m_b} + \left[ -C_F P_1 + \left(\frac{C_A}{2} - C_F\right) P_8 \right] \frac{\alpha_s}{r} + \hat{H}_{b\bar{b},\frac12} + \dots,
\end{equation}
with the color-singlet and color-octet projectors 
\begin{equation}
 (P_1)_{abcd} = \frac{1}{N_c} \, \delta_{bc}\delta_{da}, \hspace{2cm} (P_8)_{abcd} = 2 \, T_{bc}^AT_{da}^A,
\end{equation}
where the color indices are assigned in the same way as in the potential term 
in~\eqref{eq:LPNRQCD}. The LO Hamiltonian $\hat{H}_{b\bar{b},0}$ is of the order $m_b v^2$. 
The non-perturbative dynamics at the scale $\LQCD$ are described by the Hamiltonian 
$\hat{H}_\text{np}$ which is of the order $\LQCD$. The leading interaction between the 
bottomonium and non-perturbative sector is given by the chromoelectric dipole term 
\begin{equation}
 \hat{H}_D = -\frac{g_s}{2}\,\xi^A\,\mathbf{x}\cdot\mathbf{E}^{\text{np}, A}(0,\mathbf{0}),
\end{equation}
with $\xi_{abcd}^A=T_{ab}^A\delta_{cd}+\delta_{ab}T_{cd}^A$ when the color indices are 
again assigned in the same way as in the potential term in~\eqref{eq:LPNRQCD}, which 
is of the order $\LQCD^2/(m_b v)$. Assuming $\LQCD\ll m_b v^2$ the interaction 
$\hat{H}_D$ and the non-perturbative Hamiltonian $\hat{H}_\text{np}$ can therefore 
both be treated as perturbations and the physical state 
\begin{equation}
 \hm{|}\hspace{-0.12cm}\hm{|}\mathbf{0}\hm{\rangle}\hspace{-0.2cm}\hm{\rangle} \equiv |\mathbf{0}\rangle_{b\bar{b}} \otimes |0\rangle_\text{np}
\end{equation}
factorizes into the product of a bottom-antibottom state $|\mathbf{0}\rangle_{b\bar{b}}$ 
at zero spatial separation and the non-perturbative vacuum state $|0\rangle_\text{np}$. 
The expansion of the Green function~\eqref{eq:GE_def} in powers of $\LQCD$ then takes 
the form 
\begin{equation}
\begin{aligned}
 G(E) = &\, \langle\mathbf{0}|\hat{G}_{b\bar{b}}(E)|\mathbf{0}\rangle_{b\bar{b}} \, \langle0|0\rangle_\text{np} \,+  
  \sum\limits_{n=0}^\infty \Bigg[\langle\mathbf{0}|\hat{G}_{b\bar{b}}(E) \, \xi^A\hat{x}^i \left[\hat{G}_{b\bar{b}}(E)\right]^{1+2n} \xi^B\hat{x}^j \, \hat{G}_{b\bar{b}}(E)|\mathbf{0}\rangle_{b\bar{b}} \\
 & \times\langle0|\frac{g_s^2}{4}\left(E^\text{np}\right)_i^A \left[\hat{H}_\text{np}\right]^{2n} \left(E^\text{np}\right)_j^B|0\rangle_\text{np} \Bigg] + \dots \\
 = &\, \langle\mathbf{0}|\hat{G}_{b\bar{b}}^{(1)}(E)|\mathbf{0}\rangle_{b\bar{b}} \, + 
 \sum\limits_{n=0}^\infty \langle\mathbf{0}|\hat{G}_{b\bar{b}}^{(1)}(E) \hat{x}^i \left[\hat{G}_{b\bar{b}}^{(8)}(E)\right]^{1+2n} \hat{x}^i \, \hat{G}_{b\bar{b}}^{(1)}(E)|\mathbf{0}\rangle_{b\bar{b}} \; O_n\, + \dots\,,
\end{aligned}
\label{eq:GE_expansion}
\end{equation}
where
\begin{equation}
 \hat{G}_{b\bar{b}}(E) = \hat{G}_{b\bar{b}}^{(1)}(E)P_1 + \hat{G}_{b\bar{b}}^{(8)}(E)P_8 = \left(\hat{H}_{b\bar{b}}-E-i0\right)^{-1}
\end{equation}
is the perturbative part of the Green function and we adopted the notation of~\cite{Pineda:1996uk}: 
\begin{equation}
 O_n = \langle0|\frac{g_s^2}{18}\left(E^\text{np}\right)_i^A\left[\hat{H}_\text{np}\right]^{2n}\left(E^\text{np}\right)_i^A|0\rangle_\text{np}\,. 
 \end{equation}
% \begin{equation}
% \begin{aligned}
%  O_n = & \, \langle0|\frac{g_s^2}{18}\left(E^\text{np}\right)_i^A\left[\hat{H}_\text{np}\right]^{2n}\left(E^\text{np}\right)_i^A|0\rangle_\text{np}\\
%      = & \, -\frac{g_s^2}{54}\,v^{\beta_0}\dots v^{\beta_n}v^{\alpha_0}\dots v^{\alpha_n}\,\langle0|\text{Tr}\Big([D_{\beta_1}(0),[\dots[D_{\beta_n}(0),G_{\beta_0\mu}(0)]\dots]] \\
%        & \hspace{4.5cm} \times [D_{\alpha_1}(0),[\dots[D_{\alpha_n}(0),G_{\alpha_0}^\mu(0)]\dots]]\Big)|0\rangle_\text{np}. 
% \end{aligned}
% \end{equation}
%
The properties $\hat{H}_\text{np} |0\rangle_\text{np} = 0$ and 
$\langle0|g_s\left(E^\text{np}\right)_i^A|0\rangle_\text{np}=0$ have been used to remove 
insertions of $\hat{H}_\text{np}$ that are not in between insertions of $\hat{H}_D$ and 
single insertions of $\hat{H}_D$. Terms with an odd number of $\hat{H}_\text{np}$ 
insertions between the two $\hat{H}_D$ insertions vanish, because they can be related to 
the vacuum expectation values of operators with odd numbers of Lorentz indices by using 
Lorentz invariance, see~\cite{Voloshin:1978hc}. 

\begin{figure}
\centering
\includegraphics[height=2.7cm]{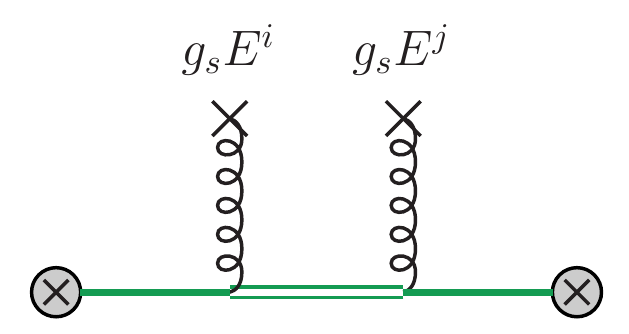}\hspace{0.5cm}\includegraphics[height=2.7cm]{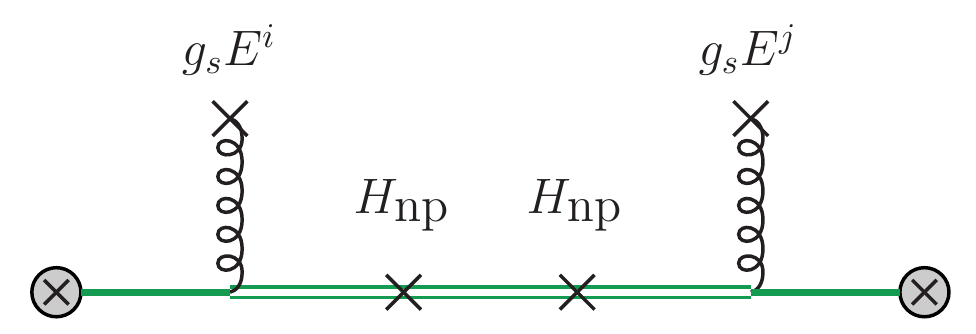}
\caption{Leading dimension four and six condensate contributions to the Green function. 
The single and double lines denote the LO color-singlet and color-octet Green functions, 
respectively. Higher-dimensional corrections are obtained by inserting additional pairs 
of the non-perturbative Hamiltonian $\hat{H}_\text{np}$ in between the two insertions of the 
chromoelectric dipole $\hat{H}_D$.}
\label{fig:dim46LO}
\end{figure}

The first term in~\eqref{eq:GE_expansion} is the purely perturbative part. The sum contains 
the leading non-perturbative contributions, which are proportional to vacuum expectation values 
of operators of even dimensions and are suppressed by $v^2(\LQCD/(m_b v^2))^{4,6,8,\dots}$ 
with respect to the perturbative expression. The contributions of dimension four and six are 
shown in Figure~\ref{fig:dim46LO}. The extra suppression factor $v^2$ is present 
because terms without at least two insertions of $\hat{H}_D$ vanish. The dimension-four 
correction contains the gluon condensate $\langle\frac{\alpha_s}{\pi}G^2\rangle$ and has been studied 
in~\cite{Voloshin:1978hc,Voloshin:1979uv,Leutwyler:1980tn,Voloshin:1995sf,Beneke:2014qea}. 
The dimension-six correction to the masses and leptonic decay rates of the $\Upsilon(NS)$ 
resonances has been calculated in~\cite{Pineda:1996uk}. 

Due to the extra suppression factor $v^2$, smallness of the dimension-four contribution is 
not sufficient to demonstrate the convergence of the expansion in $\LQCD/(m_b v^2)$ and 
the calculation of higher-order condensate corrections is necessary to gain more insight. 
We compute the leading corrections up to dimension eight in Section~\ref{sec:LO_condensates}. 
The NLO potential corrections to the dimension-four condensate contribution are determined 
in Section~\ref{sec:dim4NLO}. The size of the condensate corrections to observables in 
the $\Upsilon$ system is discussed in Section~\ref{sec:pheno}. 
We conclude in Section~\ref{sec:conclusions}.

%%%%%%%%%%%%%%%%%%%%%%%%%%%%%%%%%%%%%%%%%%%%%%%%%%%%%%%%%%%%%%%%%%%%%%%%%%%%%%%%%%%%%%%%%%
%%%%%%%%%%%%%%%%%%%%%%%%%%%%%%%     LO condensates      %%%%%%%%%%%%%%%%%%%%%%%%%%%%%%%%%%
%%%%%%%%%%%%%%%%%%%%%%%%%%%%%%%%%%%%%%%%%%%%%%%%%%%%%%%%%%%%%%%%%%%%%%%%%%%%%%%%%%%%%%%%%%

\section{Leading order condensate corrections of dimensions four, six and eight\label{sec:LO_condensates}}

The leading order condensate corrections are finite and can be computed in four dimensions 
in position space. Inserting spatial integrations, the dimension-four contribution 
in~\eqref{eq:GE_expansion} takes the form 
\begin{equation}
 \delta_{\LQCD^4}^{(0)}G(E)=O_0
\int d^3\mathbf{r}_1\int d^3\mathbf{r}_2
\left(\mathbf{r}_1\cdot\mathbf{r}_2\right)
G_0^{(1)}(0,\mathbf{r}_1;E)G_0^{(8)}(\mathbf{r}_1,\mathbf{r}_2;E)
G_0^{(1)}(\mathbf{r}_2,0;E).
\label{eq:deltaG2G_int}
\end{equation}
The integrals can be evaluated using the known representations of the LO Green function 
$G_0^{(1,8)}$, where the superscript indicates whether the bottomonium state is 
in a color singlet $(1)$ or octet $(8)$ configuration. 
It is convenient to decompose the Green function in terms of partial waves 
\begin{equation}
 G_0^{(1,8)}(\mathbf{r},\mathbf{r}';E) = \sum\limits_{l=0}^\infty\, 
    (2 l+1)\, P_l\!\left(\frac{\mathbf{r}\cdot\mathbf{r}'}{rr'}\right)\,
    G_{[l]}^{(1,8)}(r,r';E),
    \label{eq:LOGFPWD}
\end{equation}
where $l$ is the quantum number of the angular momentum of the bottom pair and $P_l(z)$ 
are the Legendre polynomials. We use an integral representation from~\cite{Wichmann:1961}, 
\begin{equation}
  \begin{aligned}
    G_{[l]}^{(1,8)}(r,r';E) =&\hspace*{0.15cm}
    \frac{m_b p}{2\pi} \frac{(2pr)^{l}(2 pr')^l}
    {\Gamma(l+1+\lambda^{(1,8)})\Gamma(l+1-\lambda^{(1,8)})}\,
    \int\limits_0^1du\int\limits_0^\infty dt \left[ut\right]^{l-\lambda^{(1,8)}}
    \\
    &\times\left[(1+t)(1-u)\right]^{l+\lambda^{(1,8)}}
    \exp\left\{-p\,[r'(1-2u)+r(1+2t)]\right\},
  \end{aligned}
  \label{eq:LOGFintegralrep}
\end{equation}
valid for $r'<r$ and a sum representation from~\cite{Voloshin:1985bd,Voloshin:1979uv}, 
\begin{equation}
    G_{[l]}^{(1,8)}(r,r';E) = \frac{m_b p}{2\pi} (2pr)^{l}(2 pr')^le^{-p(r+r')}
    \sum\limits_{s=0}^\infty\frac{s!L_s^{(2l+1)}(2pr)L_s^{(2l+1)}(2pr')}{(s+2l+1)!(s+l+1-\lambda^{(1,8)})} ,
  \label{eq:LOGFsumrep}
\end{equation}
with the Laguerre polynomials 
\begin{equation}
 L_s^{(\alpha)}(z)=\frac{e^z z^{-\alpha}}{s!}\left(\frac{d}{dz}\right)^s\left[e^{-z}z^{s+\alpha}\right].
\end{equation}
We have defined the variables 
\begin{equation}
  p=\sqrt{-m_bE}, \hspace{1cm} \lambda^{(1)}=\frac{m_b\alpha_sC_F}{2p}, \hspace{1cm} \lambda^{(8)}=\frac{m_b\alpha_s(C_F-C_A/2)}{2p}.
\label{eq:lambdaDef}
\end{equation}
In the following we take $N_c=3$ and use the variable $\lambda\equiv \lambda^{(1)} = -8\lambda^{(8)}$. 
We use the integral representation~\eqref{eq:LOGFintegralrep} for the color-singlet Green 
functions and the sum representation~\eqref{eq:LOGFsumrep} for the color-octet Green function. 
The angular integrals in \eqref{eq:deltaG2G_int} project out the S-wave component of the 
color-singlet Green functions and the P-wave component of the color-octet Green function. 
We obtain 
\begin{equation}
 \delta_{\LQCD^4}^{(0)}G(E) = R_4\,\frac{m^2\alpha_sC_F}{4\pi}\,\lambda^5\,
 \sum\limits_{s=0}^\infty\frac{s!\,H_c(s)^2}{(s+3)!(s+2+\lambda/8)},
 \label{eq:deltaG2G_sum}
\end{equation}
where 
\begin{equation}
 R_4 = \frac{O_0}{m_b^4(\alpha_sC_F)^6} = -\frac{\pi^2}{18}\;\frac{\langle\frac{\alpha_s}{\pi}G^2\rangle}{m_b^4(\alpha_sC_F)^6} \, .
 \label{eq:R4}
\end{equation}
The coefficients $H_c$ contain the remaining integrations and read 
\begin{equation}
 H_c(s) = \int\limits_0^\infty dt\left(\frac{1+t}{t}\right)^\lambda \int\limits_0^\infty d\rho \rho^4e^{-\rho(1+t)}L_s^{(3)}(\rho)
        = -\frac{(s+3)!}{s!}\lambda\frac{\Gamma(5)\Gamma(s-\lambda)}{\Gamma(5+s-\lambda)} \, .
\end{equation}
The sum in \eqref{eq:deltaG2G_sum} yields 
\begin{eqnarray}
 \delta_{\LQCD^4}^{(0)}G(E) & = & R_4\,\frac{m^2\alpha_sC_F}{4\pi}\,\lambda^5 \Bigg\{
 \frac{8 }{9 (8-9 \lambda )^2 (16-9 \lambda )^2 (16+9 \lambda)^2} \big(14155776 \nonumber\\
 & & +43024384 \lambda+212248576 \lambda ^2-136918656 \lambda ^3-607347072 \lambda ^4\nonumber\\
 & &+444623094 \lambda ^5+321157305 \lambda ^6-245939085 \lambda ^7-47534445 \lambda ^8\nonumber\\
 & &+37200870 \lambda ^9\big) + \left(\psi_1(\lambda )-\frac{\pi^2}{\sin^2(\pi\lambda)}\right)\\
 & &\times\frac{16 \lambda ^2 \left(26624-101216 \lambda ^2+109935 \lambda ^4-25515 \lambda ^6\right)}{16384-25920 \lambda ^2+6561 \lambda ^4} \nonumber\\
 & &+\frac{134217728 \lambda  \left(64-\lambda ^2\right)}{9 \left(16384-25920 \lambda ^2+6561 \lambda ^4\right)^2} \left(\psi(1-\lambda )-\psi\left(2+\frac{\lambda }{8}\right)\right)\Bigg\},\nonumber
 \label{eq:deltaG2G}
\end{eqnarray}
where $\psi$ and $\psi_1$ are the polygamma functions of order 0 and 1, respectively. 
The condensate corrections to the S-wave energy levels $E_N$ and the wave functions at 
the origin $|\psi_N(0)|^2$ can be obtained from the expansion of~\eqref{eq:deltaG2G} 
for $\lambda$ near positive integer values $N$ as described e.g. in~\cite{Beneke:2013PartII,Beneke:2013kia,Beneke:2017rdn}. 
The results are given in Appendix~\ref{sec:eANDf}. 

The same strategy can be applied for the calculation of the dimension six and eight condensate 
corrections. Again, the angular integrals project out the S-wave component of the color-singlet 
Green functions and the P-wave components of the color-octet Green functions. We find 
\begin{eqnarray}
 \delta_{\LQCD^6}^{(0)}G(E) & = & R_6\frac{m^2\alpha_sC_F}{4\pi}\lambda^9\sum\limits_{s_1=0}^\infty\sum\limits_{s_2=0}^\infty\sum\limits_{s_3=0}^\infty
 \frac{s_1!H_c(s_1)}{(s_1+3)!(s_1+2+\lambda/8)}\nonumber\\
 &   & \frac{K_c(s_1,s_2)\,s_2!\,K_c(s_2,s_3)}{(s_2+3)!(s_2+2+\lambda/8)}\frac{s_3!H_c(s_3)}{(s_3+3)!(s_3+2+\lambda/8)}, 
 \label{eq:del6Gsol}
\end{eqnarray}
and 
\begin{eqnarray}
 \delta_{\LQCD^8}^{(0)}G(E) & = & R_8\frac{m^2\alpha_sC_F}{4\pi}\lambda^{13}\sum\limits_{i=1}^5\sum\limits_{s_i=0}^\infty
 \frac{s_1!H_c(s_1)}{(s_1+3)!(s_1+2+\lambda/8)}\frac{s_5!H_c(s_5)}{(s_5+3)!(s_5+2+\lambda/8)}\nonumber\\
 &   & \hspace{-1.3cm} \times \frac{K_c(s_1,s_2)s_2!K_c(s_2,s_3)s_3!K_c(s_3,s_4)s_4!K_c(s_4,s_5)}{(s_2+3)!(s_2+2+\lambda/8)(s_3+3)!(s_3+2+\lambda/8)(s_4+3)!(s_4+2+\lambda/8)}, 
 \label{eq:del8Gsol}
\end{eqnarray}
where 
\begin{equation}
 R_6=\frac{O_1}{m_b^6(\alpha_sC_F)^{10}}, \hspace{1.5cm} R_8=\frac{O_2}{m_b^8(\alpha_sC_F)^{14}},
 \label{eq:R6_R8}
\end{equation}
and the coefficients $K_c$ read 
\begin{eqnarray}
 K_c(a,b) & = & \int\limits_0^\infty d\rho\rho^4e^{-\rho}L_a^{(3)}(\rho)L_b^{(3)}(\rho)\nonumber\\
          & = & \frac{(\text{max}(a,b)+3)!}{\text{min}(a,b)!}\left[(2a+4)\delta_{a,b}-\delta_{a,b-1}-\delta_{a-1,b}\right].
 \label{eq:Kdim6}
\end{eqnarray}
Since $K_c(a,b)$ is only non-vanishing for $|a-b|\leq1$ the multiple sums in~\eqref{eq:del6Gsol} 
and~\eqref{eq:del8Gsol} are reduced to a single sum, which can be solved in terms of polygamma 
functions. The lengthy results are available as ancillary files with the arXiv version of this 
article. The dimension six and eight contributions to the energy levels and wave functions are 
given in~Appendix~\ref{sec:eANDf}.

%%%%%%%%%%%%%%%%%%%%%%%%%%%%%%%%%%%%%%%%%%%%%%%%%%%%%%%%%%%%%%%%%%%%%%%%%%%%%%%%%%%%%%%%%%
%%%%%%%%%%%%%%%%%%%%%%%%%%%%%%%%     Dim 4 at NLO      %%%%%%%%%%%%%%%%%%%%%%%%%%%%%%%%%%%
%%%%%%%%%%%%%%%%%%%%%%%%%%%%%%%%%%%%%%%%%%%%%%%%%%%%%%%%%%%%%%%%%%%%%%%%%%%%%%%%%%%%%%%%%%

\section{Dimension four contribution at NLO: Potential contributions\label{sec:dim4NLO}}

\begin{figure}
\centering
\includegraphics[width=\textwidth]{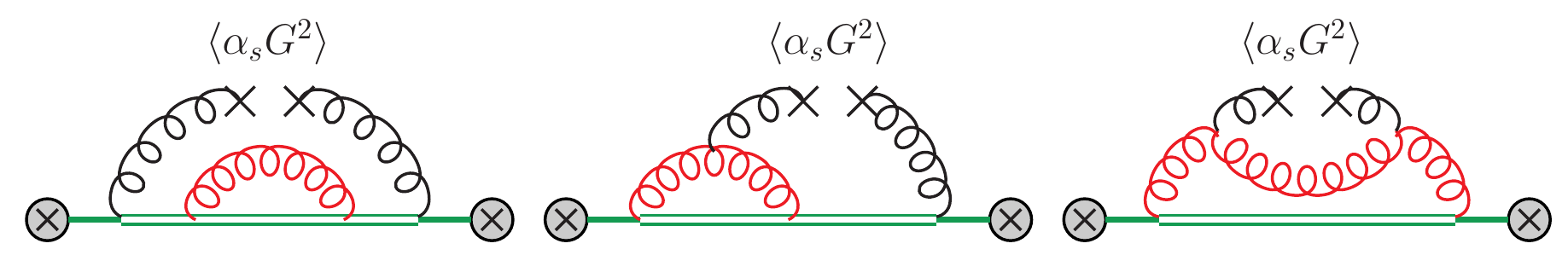}\\[0.8cm]
\includegraphics[height=2.7cm]{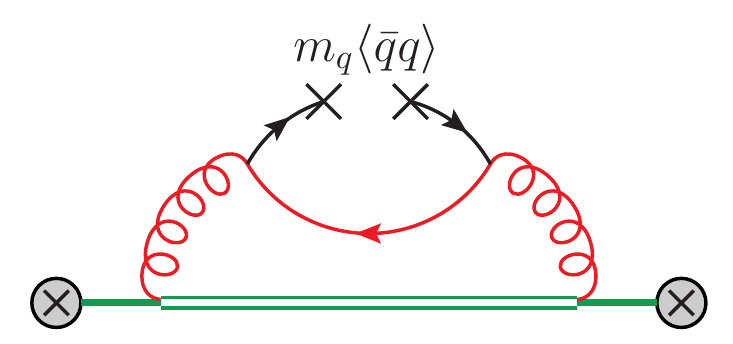}
\caption{Corrections to the dimension four condensate contribution to the 
Green function involving ultrasoft loops. Lines that carry ultrasoft momentum 
are drawn in red.}
\label{fig:dim4NLOus}
\end{figure}

The NLO corrections to the dimension-four condensate contribution involve an 
insertion of the NLO Coulomb potential as shown in Figure~\ref{fig:dim4NLOpot} and 
ultrasoft loops as shown in Figure~\ref{fig:dim4NLOus}. The upper panel of 
Figure~\ref{fig:dim4NLOus} shows the diagrams with ultrasoft gluon loops, where 
the gluon coupling to the color-octet state originates from the leading term 
$g_sA_0^\text{us}(t,\mathbf{0})$ in the multipole expansion. The equivalent 
coupling to the color singlet state vanishes because the ultrasoft gluons cannot 
resolve the spatial separation of the bottom-antibottom state and the net color 
charge vanishes in the singlet state.\footnote{See also~\cite{Beneke:2010da} 
for a more formal argument based on a field transformation.} 
The diagram in the lower panel of Figure~\ref{fig:dim4NLOus} shows the contribution 
from the light-quark condensate $\langle\bar{q}q\rangle$ with $q=u,d,s$ which is 
also counted as dimension four because, due to chirality suppression, the quark 
condensate $\langle\bar{q}q\rangle$ only appears together with one power of the 
light quark mass $m_q$ which is of the order $\LQCD$. There is a number of other 
effects that could possibly contribute at that order: 

\begin{figure}
\centering
\includegraphics[height=3.0cm]{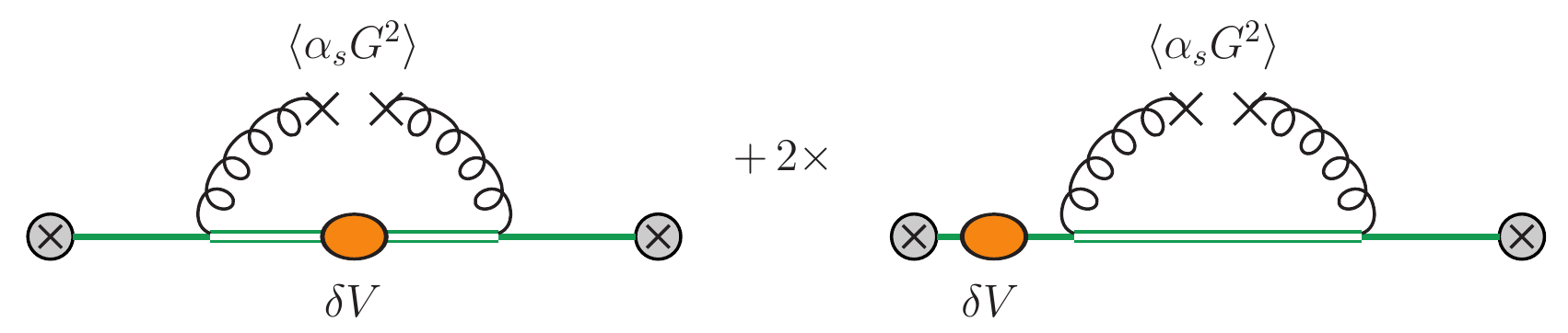}
\caption{Potential corrections to the dimension four condensate contribution to 
the Green function.}
\label{fig:dim4NLOpot}
\end{figure}

\begin{itemize}
 \item An $\alpha_s$ correction to the Wilson coefficient of the 
 chromoelectric dipole operator~\eqref{eq:Lnonpert}. The Wilson 
 coefficient was found to be trivial up to $\mathcal{O}(\alpha_s^2)$ 
 in~\cite{Brambilla:2006wp}.
 \item $\mathcal{O}(m_b^2v^3\LQCD^2)$ terms in the 
 multipole expansion \eqref{eq:Lnonpert} of the  gluon coupling 
 to bottom quarks in the spatial components. They are identical 
 to the multipole expansion of the coupling to the  ultrasoft 
 gluon field and were determined in \cite{Brambilla:2003nt}, 
 where they are denoted as $h_{SO}^{(1,0)}$. There is no NLO 
 contribution from these terms because they either have 
 vanishing tree level Wilson coefficients and are thus suppressed 
 by an additional power of $\alpha_s\sim v$ or involve the 
 chromomagnetic instead of the chromoelectric field, which only 
 yields a vanishing condensate $\Braket{0|E_i^AB_j^B|0}=0$ at NLO. 
 \item Contrary to the ultrasoft gluon-bottom coupling, the interactions 
 of the non-perturbative gluon field must also be multipole expanded in 
 the time component. The expansion of the $A^0$  component is trivial 
 due to our gauge choice~\eqref{eq:FockSchwingerGauge} and already the 
 linear term $t(\partial^0A^i)(0,\mathbf{0})$ in the expansion of the 
 spatial component is only relevant at higher powers. 
 Thus, no contributions of this type need to be considered at NLO. 
\end{itemize}
The potential corrections are determined below, whereas the 
ultrasoft contribution is postponed to future work. 
The NLO correction to the Coulomb potential is given by 
\begin{equation}
 \delta^{(1)}V^{(1,8)}(\mathbf{q}) = \frac{\alpha_s^2\,C^{(1,8)}}{\mathbf{q}^2}\,\left\{
 \left[\left(\frac{\mu^2}{\mathbf{q}^2}\right)^\epsilon-1\right]\frac{\beta_0}{\epsilon}+\left(\frac{\mu^2}{\mathbf{q}^2}\right)^\epsilon a_1(\epsilon)\right\}\,,
 \label{eq:VCoulomb}
\end{equation}
where the color factors are given by $C^{(1)} = -C_F$ and $C^{(8)} = C_A/2-C_F$ and 
\begin{eqnarray}
 a_1(\epsilon) & = & \Big(C_A[11-8\epsilon]-4T_Fn_f\Big)\frac{e^{\gamma_E\epsilon}\Gamma(1-\epsilon)\Gamma(2-\epsilon)\Gamma(\epsilon)}{(3-2\epsilon)\Gamma(2-2\epsilon)}-\frac{\beta_0}{\epsilon}, \nonumber\\
 \beta_0 & = & \frac{11C_A}{3}-\frac{4T_F n_f}{3},
\end{eqnarray}
where $n_f$ is the number of massless quarks. Denoting the contribution from 
the left (right) diagram in Figure~\ref{fig:dim4NLOpot} by DVD (DDV), we find 
\begin{eqnarray}
 \label{eq:delG41}
 \delta_{\LQCD^4}^{(1),\text{ pot}}G(E) & = &  \delta_{\LQCD^4}^{(1),\;\text{DVD}}G(E) + 2 \, \delta_{\LQCD^4}^{(1),\,\text{DDV}}G(E) \\
                          & = & -O_0\,\alpha_s^2\,\left[a_1+\beta_0\frac{d}{du}\right]\nonumber\\
                          &   & \left[\left(\frac{C_A}{2}-C_F\right)I_{\LQCD^4}[D,1+u,D] + \,2\,(-C_F)\,I_{\LQCD^4}[D,D,1+u]\right]_{u=0}.\nonumber
\end{eqnarray}
The first triple insertion function takes the form 
\begin{eqnarray}
 I_{\LQCD^4}[D,1+u,D] & = & \frac{\mu^{2u}}{4\pi\Gamma(1+2u)\cos(\pi u)}\int d^3\mathbf{r}_1\int d^3\mathbf{r}_2\int d^3\mathbf{r}_3\,(\mathbf{r}_1\cdot\mathbf{r}_3)\nonumber\\
                                 &   & \times G_0^{(1)}(0,\mathbf{r}_1;E)G_0^{(8)}(\mathbf{r}_1,\mathbf{r}_2;E)r_2^{-1+2u}G_0^{(8)}(\mathbf{r}_2,\mathbf{r}_3;E)G_0^{(1)}(\mathbf{r}_3,0;E)\nonumber\\
    & = & \frac{\lambda^6}{(4\pi)^2 m_b^2(\alpha_sC_F)^6}\sum\limits_{s_1=0}^\infty\sum\limits_{s_2=0}^\infty\frac{s_1!H_c(s_1)}{(s_1+3)!(s_1+2+\lambda/8)}\nonumber\\
    &   & \times K_V(u,s_1,s_2) \, \frac{s_2!H_c(s_2)}{(s_2+3)!(s_2+2+\lambda/8)},
    \label{eq:IGVG}
\end{eqnarray}
where 
\begin{equation}
 K_V(u,s_1,s_2)=\frac{(\mu/2p)^{2u}}{\Gamma(1+2u)\cos(\pi u)}\int\limits_0^\infty d\rho\rho^{3+2u}e^{-\rho}L_{s_1}^{(3)}(\rho)L_{s_2}^{(3)}(\rho). 
 \label{eq:KV_def}
\end{equation}
The full $u$-dependence of~\eqref{eq:KV_def} is not needed here. 
To evaluate~\eqref{eq:delG41} we only need the value and the 
first derivative at $u=0$. We obtain 
\begin{equation}
 K_V^{(0)}(s_1,s_2)\equiv K_V(0,s_1,s_2)=\frac{(s_1+3)!}{s_1!}\delta_{s_1s_2}.
\end{equation}
The derivative of~\eqref{eq:KV_def} at zero can be solved by applying the 
methods used for the Coulomb triple insertion in~\cite{Beneke:2013PartII}. 
This yields 
\begin{equation}
 K_V^{(1)}(s_1,s_2)\equiv \frac{d}{du}K_V(u,s_1,s_2)|_{u=0}=2\left[(L_\lambda+\gamma_E)\frac{(s_1+3)!}{s_1!}\delta_{s_1s_2}+k_V(s_1,s_2)\right],
\end{equation}
where $L_\lambda=\ln(\lambda\mu/(m_b\alpha_sC_F))$ and 
\begin{equation}
 k_V(s_1,s_2)=\begin{cases}
               11+12s_1+3s_1^2+\frac{(s_1+3)!}{s_1!}\psi(1+s_1),& s_1=s_2\\
               -\frac{(\text{min}(s_1,s_2)+3)!}{\text{min}(s_1,s_2)!|s_1-s_2|},& \text{else.}
              \end{cases}
\end{equation}
The second triple-insertion function yields 
\begin{eqnarray}
 I_{\LQCD^4}[D,D,1+u] & = & \frac{\mu^{2u}}{4\pi\Gamma(1+2u)\cos(\pi u)}\int d^3\mathbf{r}_1\int d^3\mathbf{r}_2\int d^3\mathbf{r}_3\,(\mathbf{r}_1\cdot\mathbf{r}_2)\nonumber\\
                      &   & \times G_0^{(1)}(0,\mathbf{r}_1;E)G_0^{(8)}(\mathbf{r}_1,\mathbf{r}_2;E)G_0^{(1)}(\mathbf{r}_2,\mathbf{r}_3;E)r_3^{-1+2u}G_0^{(1)}(\mathbf{r}_3,0;E)\nonumber\\
    & = & \frac{\lambda^6}{(4\pi)^2 m_b^2(\alpha_sC_F)^6}\sum\limits_{s_1=0}^\infty\sum\limits_{s_2=0}^\infty\frac{s_1!H_c(s_1)}{(s_1+3)!(s_1+2+\lambda/8)}\nonumber\\
    &   & \times K_D(s_1,s_2)\,\frac{H(u,s_2+1)}{(s_2+1)(s_2+1-\lambda)},
    \label{eq:IGGV}
\end{eqnarray}
where 
\begin{eqnarray}
 K_D(s_1,s_2) & = & \int\limits_0^\infty d\rho\rho^4e^{-\rho}L_{s_1}^{(3)}(\rho)L_{s_2}^{(1)}(\rho)\nonumber\\
              & = & \begin{cases}
                    (-1)^{s_1+s_2}4!\frac{(s_1+3)!}{s_2!(s_1+3-s_2)!}\frac{(s_2+1)!}{s_1!(s_2+1-s_1)!},& -1\leq s_2-s_1\leq3\\
                    0,& \text{else,}
                    \end{cases}
\end{eqnarray}
and $H(u,k)$ is defined as in \cite{Beneke:2013PartII}. Also here, 
we only need the  value and the first derivative at $u=0$: 
\begin{eqnarray}
 H^{(0)}(k) & \equiv & H(0,k) = \frac{k}{k-\lambda},\\
 H^{(1)}(k) & \equiv & \frac{\p}{\p u}H(u,k)|_{u=0} \nonumber\\
 &=&\frac{2k}{k-\lambda}\left[L_\lambda-\gamma_E-\psi(k-\lambda)+\frac{\lambda}{k}\left(\psi(1-\lambda)-\psi(k+1-\lambda)\right)\right]. 
\end{eqnarray}
The infinite sums in \eqref{eq:IGVG} and \eqref{eq:IGGV} converge 
quickly and can be truncated with negligible uncertainty at 
$s_i\sim30$ for the numerical evaluation of the Green function.  
The contributions to the energy levels and wave functions 
from the potential corrections can be extracted by expanding 
\eqref{eq:IGVG} and \eqref{eq:IGGV} for $\lambda$ near positive 
integer values $N$. The results are given in Appendix~\ref{sec:eANDf}.

%%%%%%%%%%%%%%%%%%%%%%%%%%%%%%%%%%%%%%%%%%%%%%%%%%%%%%%%%%%%%%%%%%%%%%%%%%%%%%%%%%%%%%%%%%
%%%%%%%%%%%%%%%%%%%%%%%%%%%%%%%%     Phenomenology      %%%%%%%%%%%%%%%%%%%%%%%%%%%%%%%%%%
%%%%%%%%%%%%%%%%%%%%%%%%%%%%%%%%%%%%%%%%%%%%%%%%%%%%%%%%%%%%%%%%%%%%%%%%%%%%%%%%%%%%%%%%%%

\section{Phenomenology of condensate corrections\label{sec:pheno}}

The size of non-perturbative corrections to the moments and to the properties 
of the $\Upsilon$ resonances has been strongly disputed for various reasons. 
First, the assumption $\LQCD\ll m_bv^2$ is questionable and is certainly only 
valid for a limited number of observables in the $\Upsilon$ system. 
Here, we perform an unbiased analysis of the expansion in terms of local condensates 
and assess the validity based on its convergence. The breakdown of this expansion 
is a clear indication that the above assumption is inappropriate. 

Furthermore, the numerical values of the local condensates are very uncertain. 
The condensate $O_0$ is proportional to the gluon condensate and we will use the 
standard value $\langle\frac{\alpha_s}{\pi}G^2\rangle^\text{SVZ}=0.012\text{\,GeV}^4$ 
from~\cite{Shifman:1978by} below, unless indicated otherwise. We note however, 
that significantly larger values have also been obtained in the literature, see 
e.g.~\cite{Bali:2014sja,Dominguez:2014pga}. Clearly, the situation is even more 
uncertain for the higher-dimensional condensates. Since our main objective is the 
assessment of the convergence properties, we rely on naive rescaling  
\begin{equation}
 O_0^\text{SVZ} = -(285\text{\,MeV})^4, \hspace{1cm} O_1^\text{naive} = (285\text{\,MeV})^6, \hspace{1cm} O_2^\text{naive} = -(285\text{\,MeV})^8. 
 \label{eq:values_condensates}
\end{equation}
The value of $O_0$ is scale independent since the gluon condensate 
$\langle\frac{\alpha_s}{\pi}G^2\rangle$ is not renormalized. We neglect the 
scale dependence of the higher-dimensional condensates which is very weak 
compared to that of the coefficients which contain large powers of $\alpha_s$. 
The estimate $O_1^\text{naive}$ is in good agreement with the result of~\cite{Pineda:1996uk}, 
where an expression for $O_1$ in terms of the dimension-six gluon condensate 
$\langle G^3\rangle$ and the quark-condensate $\langle \bar{q}q\rangle$ has been 
derived based on the factorization hypothesis. The analysis of~\cite{Pineda:1996uk} 
also shows that $O_1$ is only weakly scale dependent. 

In addition, the corrections to the masses and leptonic decay rate depend 
strongly on the renormalization scale, because large powers of $\alpha_s$ appear 
in the ratios~\eqref{eq:R4} and \eqref{eq:R6_R8}. The fact that different powers 
of $\alpha_s$ appear in the contributions of different dimensions also complicates 
the assessment of the convergence and different conclusions have been drawn based 
on different scale choices. We distinguish the scale $\mu_c$, used in the condensate 
corrections, from the renormalization scale $\mu$ in the perturbative contribution. 
The main motivation for the calculation of the potential corrections to the 
dimension-four contribution has been to gain more insight into the appropriate 
scale choice for $\mu_c$ by considering the convergence of the perturbative series. 
We note that the potential corrections contain all logarithms $\ln(\mu_c)$ that are 
required to cancel the $\mu_c$ dependence of the dimension-four contribution at NLO. 
The ultrasoft correction must therefore be free of logarithms $\ln(\mu_c)$ and is 
less scale dependent, which justifies performing this analysis based on incomplete 
NLO corrections. Scales below 0.8 GeV are not considered below, because the value 
of $\alpha_s$ and perturbation theory in general become unreliable in this regime.

\subsection{\boldmath The $\Upsilon(1S)$ mass\label{sec:MUps1S}}

\begin{figure}
 \begin{center}
    \includegraphics[height=7.8cm]{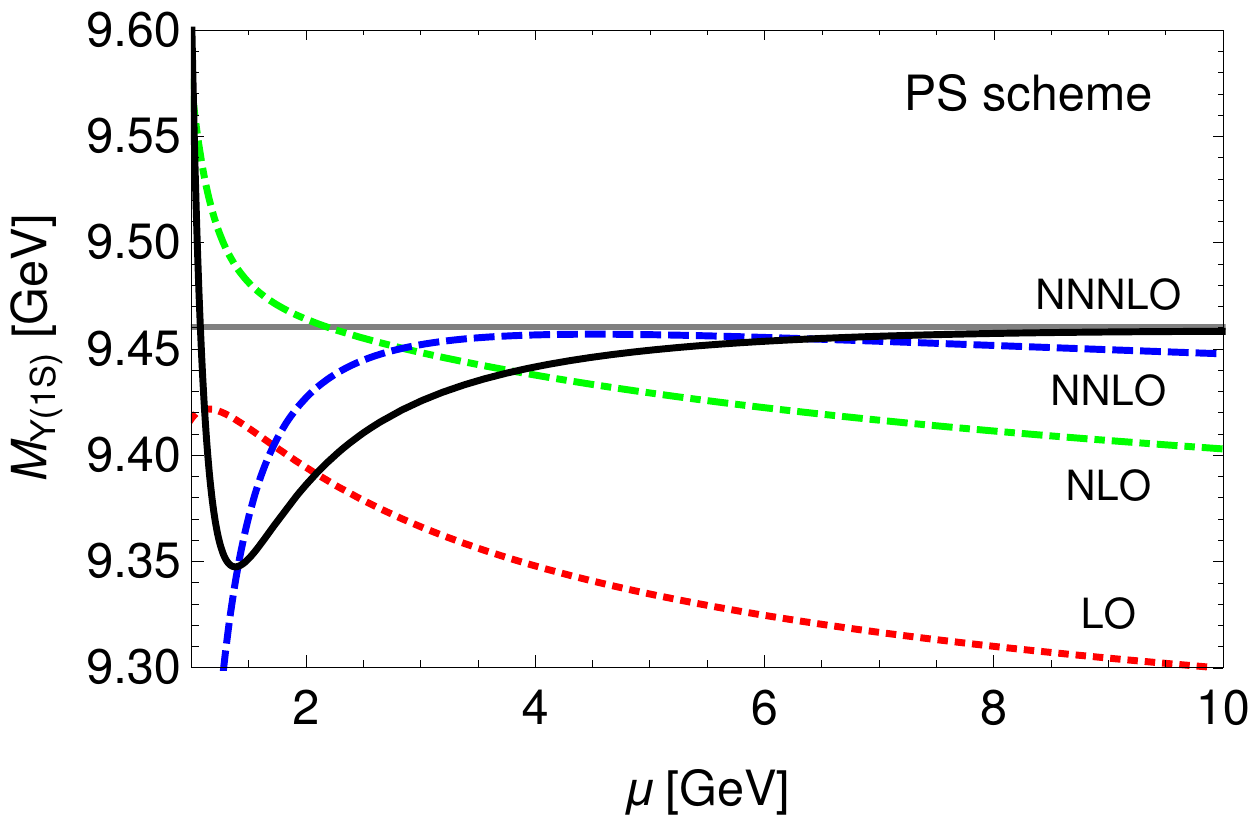}\\[0.8cm]
    \includegraphics[height=7.8cm]{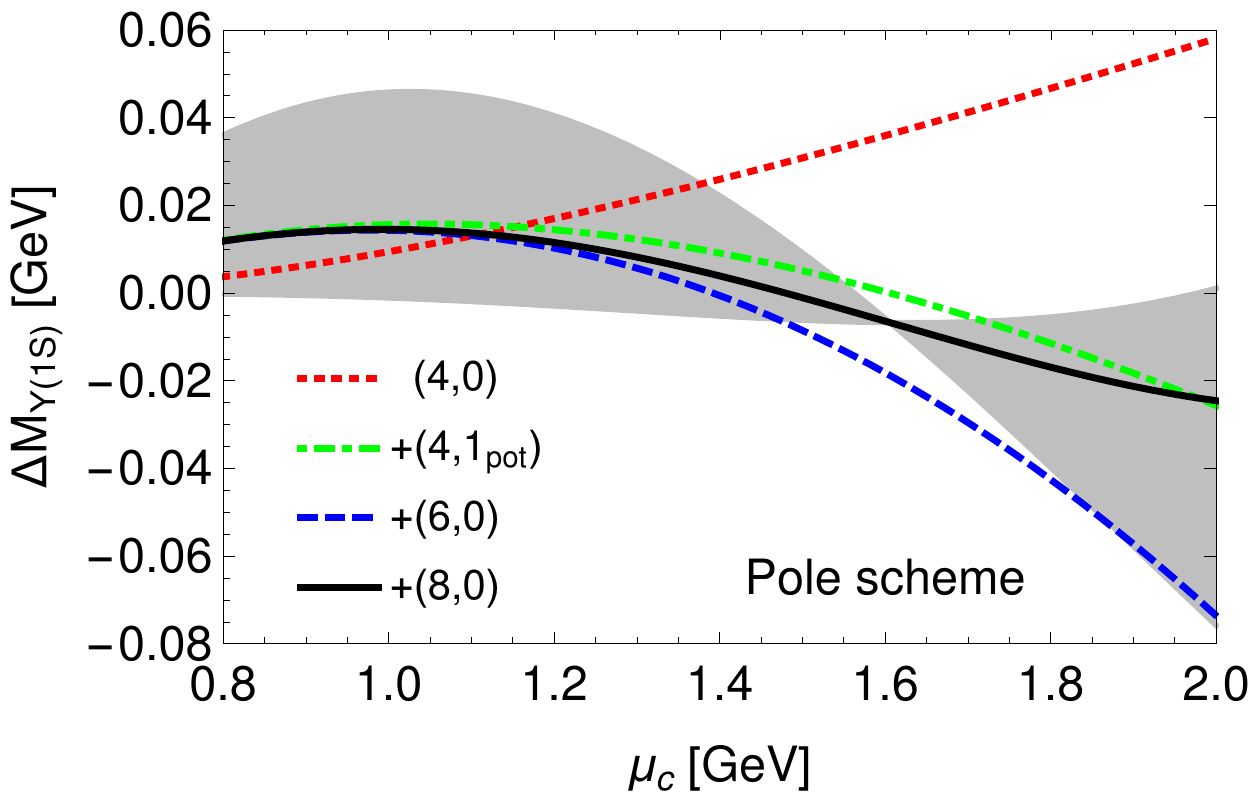}
  \caption{The top panel shows the perturbative contribution to the mass of the 
  $\Upsilon(1S)$ resonance. The curves in the bottom panel show the effects of 
  cumulatively adding the condensate contributions $(i,j)$ where $i$ denotes the 
  dimension and $j$ the order in perturbation theory. The gray band is spanned 
  by variation of $O_0$ by factors of 0 and 3, while $O_1$ and $O_2$ are unchanged.}
  \label{fig:MUps1S}
 \end{center}
\end{figure}

First, we briefly review the status of the purely perturbative prediction for the 
mass of the $\Upsilon(1S)$ resonance. We use \texttt{QQbar\_Threshold}~\cite{Beneke:2016kkb,Beneke:2017rdn} 
in the PS mass scheme~\cite{Beneke:1998rk} with the input value 
$m_b^\text{PS} = 4.532_{-0.039}^{+0.013}\text{\,GeV}$ from~\cite{Beneke:2014pta,Beneke:2016oox}. 
The effects of a non-zero charm-quark mass are included up to NNLO~\cite{Beneke:2014pta} 
using the mass $\overline{m}_c(3\text{\,GeV})=993\text{\,MeV}$ from~\cite{Chetyrkin:2009fv,Chetyrkin:2017lif}.
The default values of~\texttt{QQbar\_Threshold} are taken for the strong coupling 
$\alpha_s(m_Z) = 0.1184\pm0.0010$ and all other parameters, and QED corrections are 
taken into account with NNLO accuracy. 
The result is shown in the top panel of Figure~\ref{fig:MUps1S}. We observe that 
the convergence is best for scales that are considerably larger than the soft scale 
$\mu\sim m_b\alpha_s(\mu)C_F$. This has motivated the authors of~\cite{Kiyo:2015ufa} 
to choose a central scale of 5.35\,GeV, which is significantly larger than that of 
\cite{Ayala:2014yxa,Ayala:2016sdn,Mateu:2017hlz} (2.5 and 1.9\,GeV, respectively) 
and leads to a much smaller estimate for the perturbative uncertainty. Here, we 
choose a central scale of 3\,GeV such that a variation by factors 1/2 and 2 covers 
the choices of \cite{Ayala:2014yxa,Ayala:2016sdn,Kiyo:2015ufa,Mateu:2017hlz}. The 
perturbative expansion takes the form 
\begin{equation}
 M_{\Upsilon(1S)}^\text{pert}\,(3\text{\,GeV}) = (9\,366 + 82 + 4 - 27)\text{\,MeV}.
\end{equation}
In addition to the perturbative uncertainty from scale variation, we also take into 
account the parametric uncertainty from the bottom-quark PS mass and we use the size 
of the charm-quark mass effects up to NNLO as an estimate for the missing NNNLO correction. 
The parametric uncertainty from the strong coupling is small in the PS mass scheme 
and is neglected. 

The condensate corrections with the values of~\eqref{eq:values_condensates} are shown 
in the lower panel of Figure~\ref{fig:MUps1S}. At the considered orders, the mass 
scheme is ambiguous and we use the one-loop pole mass in the condensate contribution. 
In the PS scheme, the condensate contributions are slightly enhanced and the convergence 
is slightly worsened but the overall conclusions are unchanged. We observe that the 
potential corrections to the dimension-four condensate contribution stabilize the 
behaviour under scale variation and show a clear preference for rather small scales 
around $\mu_c = 1.2\text{\,GeV}$, which we take as the central value. The condensate 
contribution takes the form 
\begin{equation}
 \Delta M_{\Upsilon(1S)}^\text{cond}\,(1.2\text{\,GeV}) = \left[(17 - 3)\,\frac{O_0}{O_0^\text{SVZ}} - 4 \,\frac{O_1}{O_1^\text{naive}} + 1 \,\frac{O_2}{O_2^\text{naive}}\right]\text{\,MeV}. 
\end{equation}
The grey band in Figure~\ref{fig:MUps1S} is obtained by varying the value of the 
condensate $O_0$ between 0\,GeV$^4$ and $3O_0^\text{SVZ}$. In our analysis 
the condensates of dimension six and eight are varied between 0\,GeV$^6$ and 
$3^{3/2}O_1^\text{naive}$, and 0\,GeV$^8$ and $3^2O_2^\text{naive}$, respectively. 
We use this variation at the central scale as an estimate for the uncertainty from 
the value of the condensates. 

The condensate expansion becomes unstable near $\mu_c\approx2\,\text{GeV}$ where the 
LO dimension four, six and eight contributions are all of the same size. The variation 
of $\mu_c$ between 0.8\,GeV and 2\,GeV yields an uncertainty of $_{-36}^{+3}$\,MeV. 
We take $\pm36$\,MeV as an estimate for the perturbative uncertainty in order to also 
account for the unknown ultrasoft NLO correction. Combining the perturbative and 
condensate contributions we find 
\begin{eqnarray}
 M_{\Upsilon(1S)} & = & 9\,437\,_{-114}^{+61}\text{ MeV}\nonumber\\
 & = & 9\,437\,_{-74}^{+28}\,(\mu)\,_{-75}^{+25}\,(m_b)\,_{-1}^{+0}\,(\alpha_s)\,\pm9\,(m_c)\nonumber\\
 & & \pm36\,(\mu_c)\,_{-14}^{+29}\,(O_0)\,_{-18}^{+4}\,(O_1)\,_{-1}^{+10}\,(O_2)\text{ MeV},
 \label{eq:MUps1S}
\end{eqnarray}
which is in good agreement with the experimental value 
$M_{\Upsilon(1S)}^\text{exp} = 9\,460.30\pm0.26$\,MeV. 

\begin{figure}[t]
 \begin{center}
    \includegraphics[height=7cm]{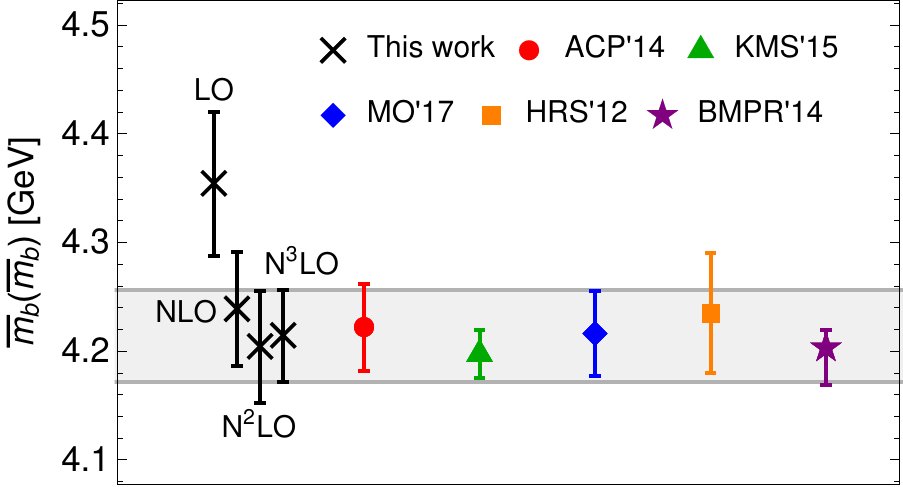}
  \caption{Comparison of our result for the bottom-quark \MS{} mass from $M_{\Upsilon(1S)}$
  with other recent results from the masses of $b\bar{b}$ bound states 
  (ACP'14~\cite{Ayala:2014yxa,Ayala:2016sdn}, KMS'15~\cite{Kiyo:2015ufa}, MO'17~\cite{Mateu:2017hlz}) 
  and non-relativistic sum rules (HRS'12~\cite{Hoang:2012us}, BMPR'14~\cite{Beneke:2014pta,Beneke:2016oox}). 
  The bottom-quark masses obtained from lower orders in pure perturbation theory, while 
  retaining all known condensate contributions, are shown as well. The order of the PS-\MS{} 
  mass relation has been correlated with the order in perturbation theory.}
  \label{fig:MbMSPlot}
 \end{center}
\end{figure}

The stable behaviour of the condensate corrections in the range 
$0.8\,\text{GeV}\lesssim\mu_c\lesssim2\,\,\text{GeV}$ facilitates the determination of the 
bottom-quark mass from the experimental value of the $\Upsilon(1S)$ mass. We obtain 
\begin{equation}
 m_b^\text{PS}(2\,\text{GeV}) = 4544\pm39\,(\text{pert.})\,_{-25}^{+22}\,(\text{non-pert.})\text{ MeV} = 4544\,_{-46}^{+44}\text{ MeV},
 \label{eq:mbPS}
\end{equation}
where we have symmetrized the uncertainty from variation of the renormalization 
scale $\mu$, by taking the maximum of the positive and negative error. The perturbative 
uncertainty is obtained by adding the errors from variation of $\mu$ and $\alpha_s$ 
as well as our estimate of higher-order charm-quark mass effects in quadrature. 
The variation of the scale $\mu_c$ and of the values of the condensates is combined 
into the non-perturbative uncertainty. The result \eqref{eq:mbPS} is converted to the 
\MS{} scheme at NNNLO~\cite{Marquard:2015qpa,Marquard:2016dcn} 
using \texttt{QQbar\_Threshold}. We distinguish the scale $\mu_m$ used in the conversion, 
which is set to $m_b^\text{PS}$, and estimate the uncertainty through variation of $\mu_m$ 
by factors of 1/2 and 2 and symmetrization as described above. We find 
\begin{equation}
 \overline{m}_b(\overline{m}_b) = 4214\pm37\,(\text{pert.})\,_{-22}^{+20}\,(\text{non-pert.})\text{ MeV} = 4214\,_{-43}^{+42}\text{ MeV}. 
 \label{eq:mbMS}
\end{equation}
The result shows good convergence and agrees with other recent determinations 
of $m_b$ from the data on the $\Upsilon$ system as shown in Figure~\ref{fig:MbMSPlot}. 
In conclusion, our analysis demonstrates that the determination of the bottom-quark 
mass from the $\Upsilon(1S)$ mass is possible with a total uncertainty of the order 
of $\pm$45\,MeV. It should however be noted that this approach to the determination 
of the bottom-quark mass is on a less sound footing theoretically than the 
extraction based on non-relativistic moments with $n\approx10$, which are discussed 
in Section~\ref{sec:moments}.

\subsection{\boldmath The $\Upsilon(2S)$ mass\label{sec:MUps2S}}

\begin{figure}
 \begin{center}
    \includegraphics[height=7.8cm]{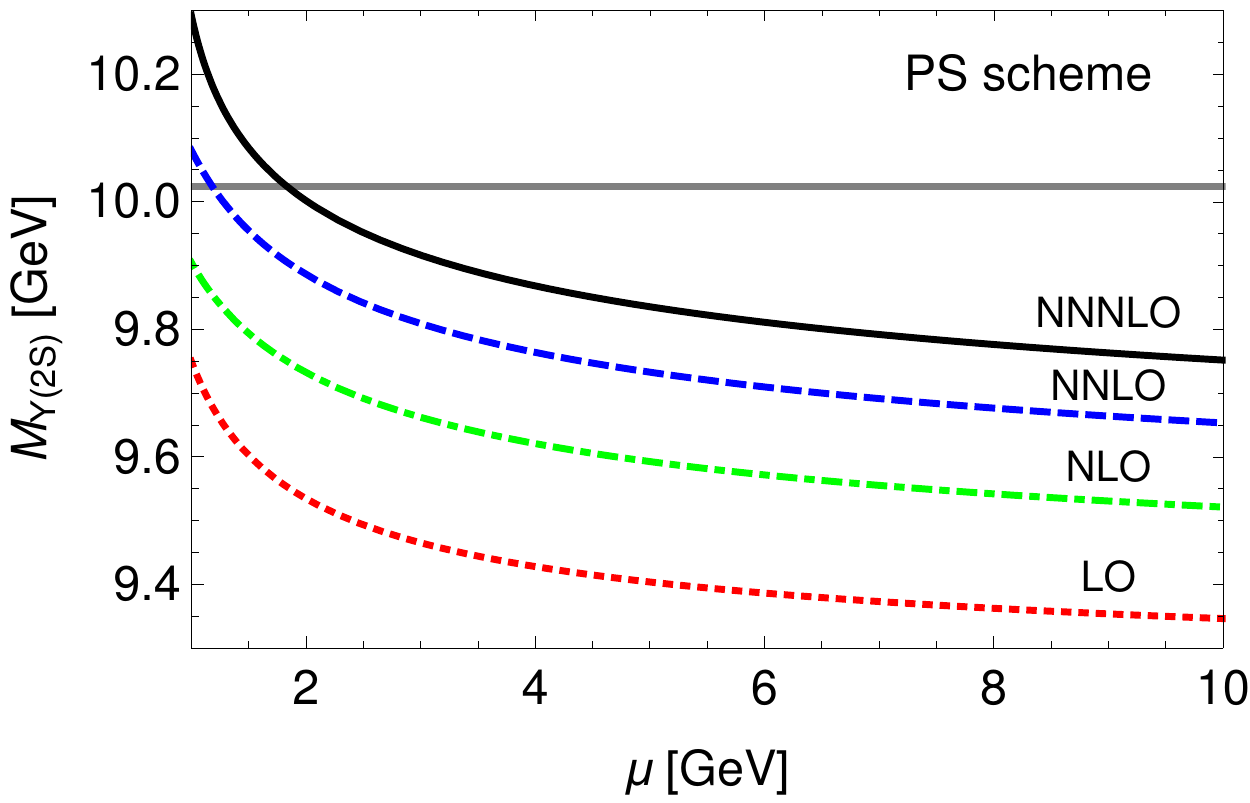}\\[0.8cm]
    \includegraphics[height=7.8cm]{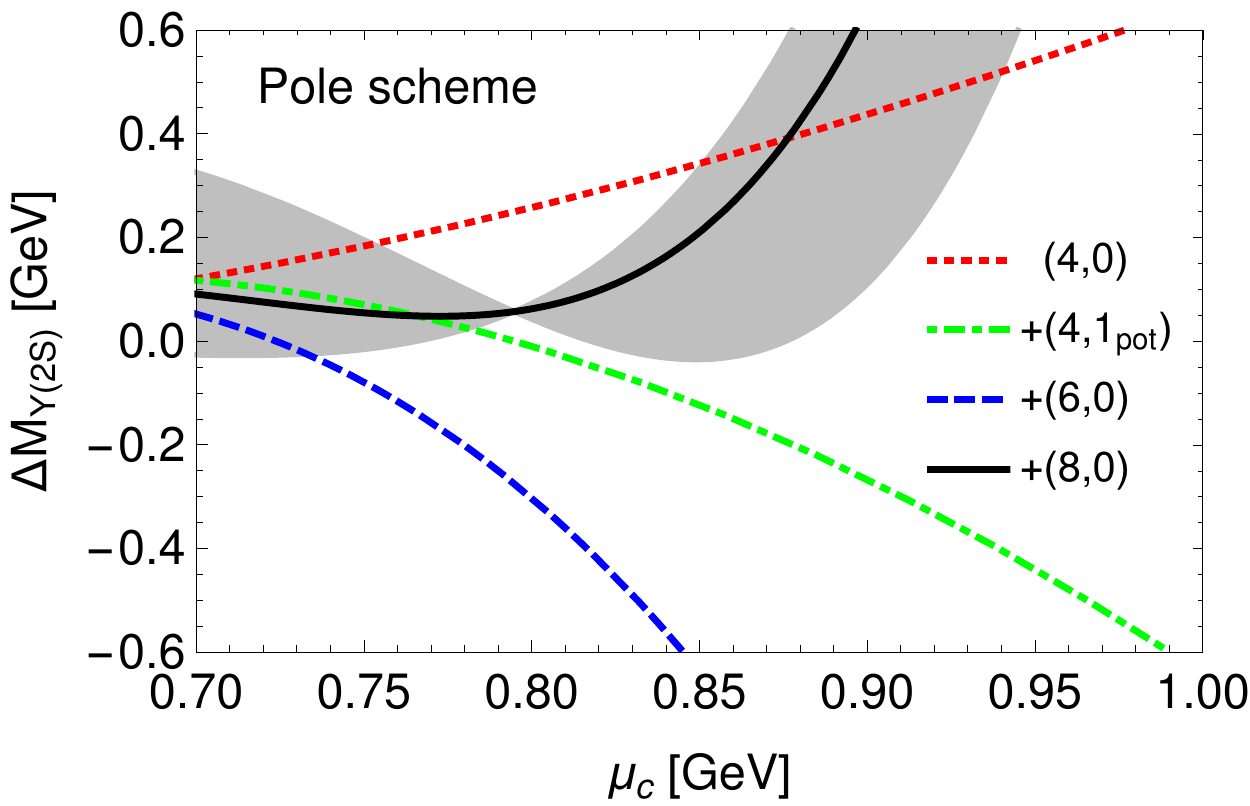}
  \caption{The top panel shows the perturbative contribution to the mass of the 
  $\Upsilon(2S)$ resonance. The curves in the bottom panel show the effects of 
  cumulatively adding the condensate contributions $(i,j)$ where $i$ denotes the 
  dimension and $j$ the order in perturbation theory. The gray band is spanned 
  by variation of $O_0$ by factors of 0 and 3, while $O_1$ and $O_2$ are unchanged.}
  \label{fig:MUps2S}
 \end{center}
\end{figure}

We repeat the above discussion for the $\Upsilon(2S)$ mass. The scale dependence 
of the perturbative result is shown in Figure~\ref{fig:MUps2S}. Since the soft 
scale is lower for the $n=2$ states, we reduce the central scale to 2\,GeV, where 
the perturbative series takes the form 
\begin{equation}
 M_{\Upsilon(2S)}^\text{pert}\,(2\text{\,GeV}) = (9\,534 + 198 + 154 + 116)\text{\,MeV}.
\end{equation}
As the plot shows, the convergence is rather slow, independently of the choice of scale. 
We also note that the charm-mass effects at NNLO are +39\,MeV and significantly larger 
than for the $\Upsilon(1S)$ mass (+8\,MeV). As we argued in~\cite{Beneke:2014pta}, the 
charm-mass effects are a measure for the IR sensitivity of an observable. Thus, the 
significantly larger value is an indication that the non-perturbative correction should 
be considerably larger and less convergent for the $\Upsilon(2S)$ mass than for the the 
$\Upsilon(1S)$ mass. 

Turning to the condensate corrections, which are shown in the lower panel of Figure~\ref{fig:MUps2S}, 
we can confirm this expectation. The expansion already breaks down for $\mu_c=0.8$\,GeV, 
where the individual contributions are 
\begin{equation}
 \Delta M_{\Upsilon(2S)}^\text{cond}\,(0.8\text{\,GeV}) = \left[(258 - 267)\,\frac{O_0}{O_0^\text{SVZ}} - 293\,\frac{O_1}{O_1^\text{naive}} + 365 \,\frac{O_2}{O_2^\text{naive}}\right]\text{\,MeV}. 
\end{equation}
At lower scales, the use of perturbation theory cannot be justified. Thus, 
while we cannot rule out the convergence of the local condensate expansion 
unambiguously due to the large uncertainties of the $O_i$, clearly no 
reliable prediction for the non-perturbative contribution can be obtained 
like this. 

A more promising approach to the $\Upsilon(2S)$ mass is to assume the 
hierarchy $\LQCD\sim m_bv^2\ll m_bv$. Then, the ultrasoft contribution takes 
the form of a non-local condensate instead of a perturbative correction~\cite{Brambilla:1999xf,Pineda:2001zq,Brambilla:2004jw}. 
This implies that the leading non-perturbative correction is of the order 
\begin{equation}
 \Delta M_{\Upsilon(2S)}^\text{non-perturbative} \sim m_b\alpha_s^2\times\left(\frac{\LQCD}{m_b\alpha_s}\right)^2\times\rho\left(\frac{\LQCD}{m_b\alpha_s^2}\right) \sim m_b\alpha_s^4,
 \label{eq:DeltaMUps2S_nonpert}
\end{equation}
which is formerly of NNLO, and the conclusion that the local condensate expansion 
breaks down is equivalent to the statement that the $\Upsilon(2S)$ system is outside 
the radius of convergence for the presently unknown function $\rho$. In this scenario, 
the perturbative NNNLO results, which contain the perturbative evaluation of the 
ultrasoft contribution cannot be used and we have to resort to the NNLO expressions. 
The result for the $\Upsilon(2S)$ mass reads 
\begin{equation}
 M_{\Upsilon(2S)} = 9\,886\,_{-122}^{+195}\,(\mu)\,_{-76}^{+25}\,(m_b)
 %\,\pm39\,(m_c) %at NNLO all mc-effects are included
 \,_{-26}^{+28}\,(\alpha_s)\pm\mathcal{O}(100)\,(\text{non-pert.})\text{ MeV},
\end{equation}
where the estimate for the non-perturbative contributions follows from the assumption 
that the function $\rho$ in \eqref{eq:DeltaMUps2S_nonpert} is of order one. Within 
the large uncertainty, the experimental value $M_{\Upsilon(2S)}^\text{exp} = 10023.26\pm0.31$\,MeV 
can be reproduced.

\subsection{\boldmath The $\Upsilon(1S)\to l^+l^-$ decay width\label{sec:Ups1S_to_ll}}

\begin{figure}
 \begin{center}
    \includegraphics[height=7.4cm]{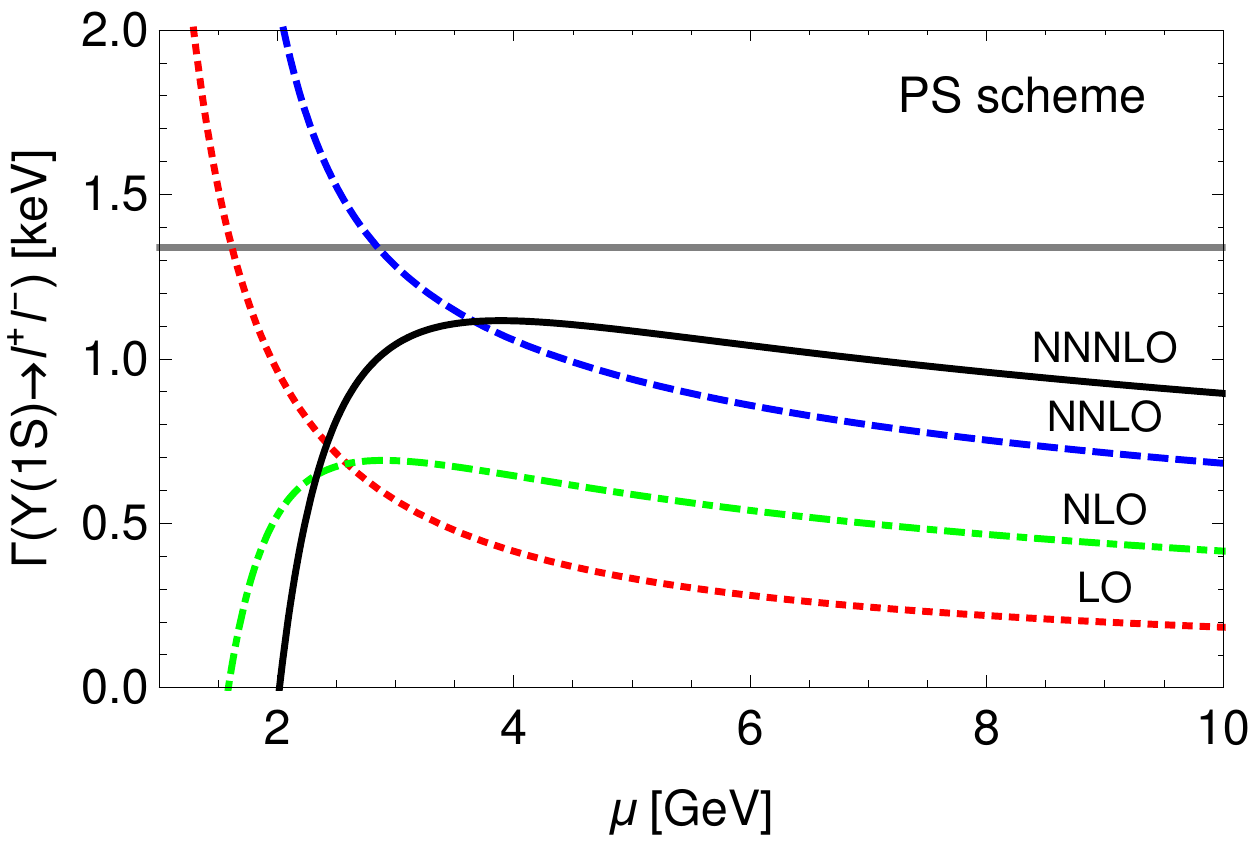}\\[0.8cm]
    \includegraphics[height=7.4cm]{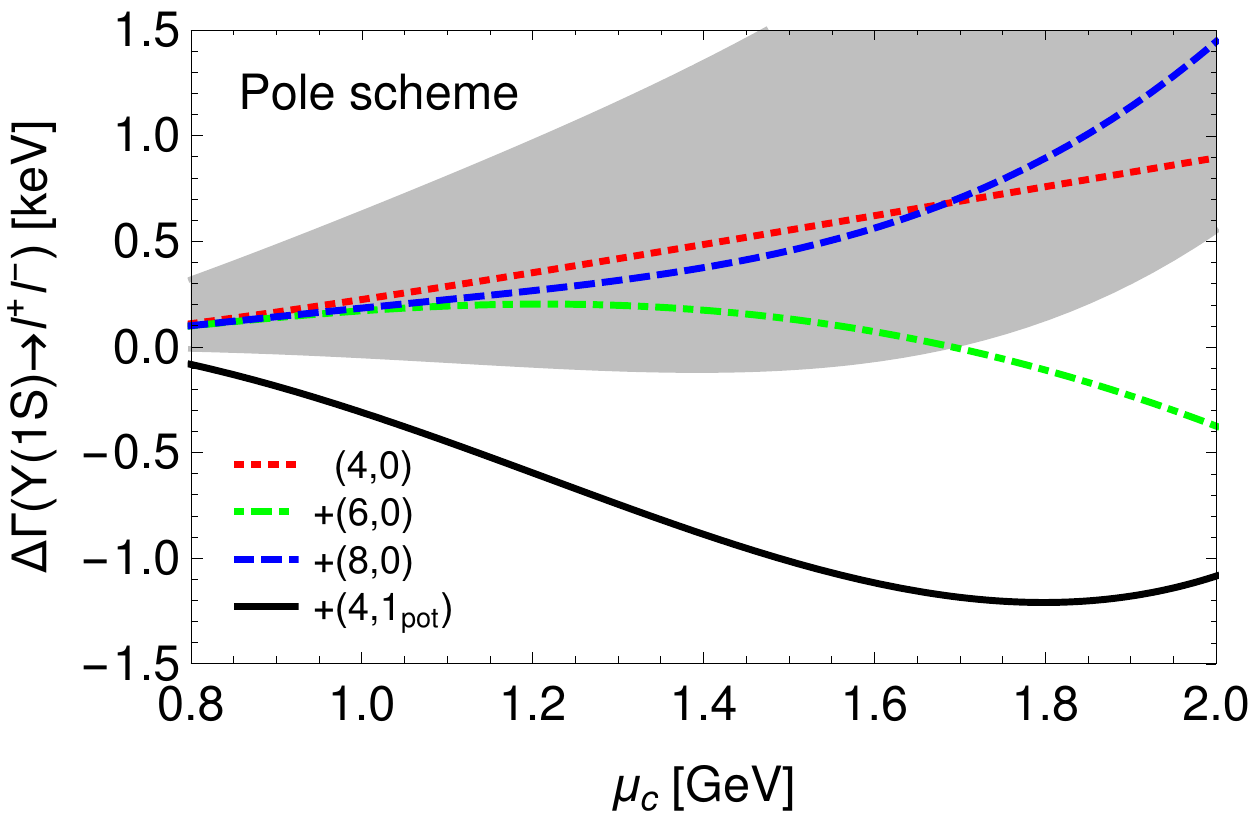}
  \caption{The top panel shows the perturbative contribution to the leptonic decay width of the 
  $\Upsilon(1S)$ resonance. The curves in the bottom panel show the effects of 
  cumulatively adding the condensate contributions $(i,j)$ where $i$ denotes the 
  dimension and $j$ the order in perturbation theory. The gray band is spanned 
  by variation of $O_0$ by factors of 0 and 3, while $O_1$ and $O_2$ are 
  unchanged. In this Figure the gray band does not contain the potential 
  corrections to the dimension-four contribution.}
  \label{fig:GammaUps1S}
 \end{center}
\end{figure}

The perturbative NNNLO result for the leptonic decay width of the $\Upsilon(1S)$ resonance 
has been obtained in~\cite{Beneke:2014qea}. Here, we repeat their analysis including 
charm-mass effects up to NNLO, which increase the leptonic width by 0.03\,keV. The scale 
dependence is shown in Figure~\ref{fig:GammaUps1S} and we adopt 3.5\,GeV as the central 
scale. The perturbative series stabilizes at NNNLO 
\begin{eqnarray}
 & & \Gamma^\text{pert}(\Upsilon(1S)\to l^+l^-)\,(3.5\,\text{GeV}) = 
 \frac{4\pi\alpha^2}{9 m_b^2}\, c_v\left[c_v - \left(c_v + \frac{d_v}{3}\right)\,\frac{E_1}{m_b}\right]\left|\psi_1(0)\right|^2 \nonumber\\
 & = & (0.48 + 0.19 + 0.47 - 0.04)\,\text{keV} = 1.11\,_{-0.22}^{+0.07}\,\text{keV}\nonumber\\
 & = & 1.11\,_{-0.21}^{+0.01}\,(\mu)\pm\,0.00(m_b)\pm\,0.04(m_c)\pm\,0.05(\alpha_s)\,\text{keV}
\end{eqnarray}
but falls short of the experimental value $\Gamma^\text{exp}(\Upsilon(1S)\to l^+l^-) = 1.340\pm0.018$\,keV
by about 20\%. Following~\cite{Beneke:2014qea} we determine the scale uncertainty from 
variation between 3 and 10\,GeV. The other input parameters are varied as above. 

The condensate contributions are shown in the lower panel of Figure~\ref{fig:GammaUps1S}. 
Using the same central scale $\mu_c = 1.2$\,GeV as for the $\Upsilon(1S)$ mass, we obtain 
\begin{equation}
 \Gamma^\text{cond}(\Upsilon(1S)\to l^+l^-)\,(1.2\text{\,GeV}) = \left[(352 - 862)\,\frac{O_0}{O_0^\text{SVZ}} - 149\,\frac{O_1}{O_1^\text{naive}} + 64 \,\frac{O_2}{O_2^\text{naive}}\right]\text{\,eV}. 
\end{equation}
Focusing first on the leading-order contributions, we see that the expansion converges and 
yields a contribution of 0.27\,keV that closes the difference between the perturbative and 
the experimental value. Compared to the $\Upsilon(1S)$ mass the expansion breaks down at a 
smaller scale around 1.6\,GeV. 

However, with the addition of the potential corrections to the dimension four contribution, 
the agreement is destroyed. The potential correction already exceeds the LO term at the 
scale 0.7\,GeV and becomes twice as large at 0.9\,GeV. This apparent breakdown of the 
perturbative series makes it impossible to give a reliable estimate of the non-perturbative 
contribution. However, it is conceivable that the large potential corrections are compensated 
by the missing ultrasoft correction, thus stabilizing the perturbative expansion of the 
dimension-four contribution. Therefore, no definite conclusions about the validity of the 
local condensate expansion for $\Gamma(\Upsilon(1S)\to l^+l^-)$ can be drawn without a 
calculation of the full NLO corrections to the dimension-four contribution.

\subsection{The non-relativistic moments\label{sec:moments}}

The moments ${\cal M}_n$ of the normalized inclusive $b\bar{b}$ production cross section 
\begin{equation}
  \label{eq:Rb_def}
  R_b(s) = \frac{\sigma(e^+e^- \to b\overline{b} + X)}{\sigma(e^+e^- \to
    \mu^+ \mu^-)}\,,
\end{equation}
in $e^+e^-$ collisions with the center-of-mass energy $s$, are defined as 
\begin{equation}
\label{eq:Mn_def}
{\cal M}_n \equiv \int_0^\infty ds\,\frac{R_b(s)}{s^{n+1}}
= - 6 \pi i \oint_{{\cal C}} ds\,\frac{\Pi_b(s)}{s^{n+1}}
= \frac{12 \pi^2}{n!} \bigg(\frac{d}{dq^2}\bigg)^n
\Pi_b\big(q^2\big)\bigg|_{q^2 = 0}\,.
\end{equation}
The normalized cross section is related to the bottom-quark contribution $\Pi_b$ to 
the photon vacuum polarization by the optical theorem $R_b(s) = 12 \pi \im \Pi_b(s+i\epsilon)$. 
The contour ${\cal C}$ must be closed around $s=0$ without crossing the branch cut 
for real $s\geq M_{\Upsilon(1S)}^2$. The perturbative contributions to the moments up to 
NNNLO have been discussed in detail in~\cite{Beneke:2014pta}. The non-perturbative corrections 
can be determined by inserting the condensate contribution to the cross section 
\begin{eqnarray}
  \Pi_b^\text{cond}(s) & = & \frac{2N_ce_b^2}{s}\,\Bigg[\delta_{\LQCD^4}^{(0)}G(E) + \left(2c_v^{(1)}\,\frac{\alpha_s}{4\pi}\,\delta_{\LQCD^4}^{(0)}G(E) + \delta_{\LQCD^4}^{(1),\text{pot}}G(E)\right) \nonumber\\
  & & \hspace{1.3cm}+ \delta_{\LQCD^6}^{(0)}G(E) + \delta_{\LQCD^8}^{(0)}G(E)+ \dots\Bigg]\,,
 \label{eq:Pib_cond}
\end{eqnarray}
where $E=\sqrt{s}-2m_b$ and $c_v^{(1)} = -8C_F$ is the hard matching coefficient of 
the vector current, into~\eqref{eq:Mn_def}. Following the discussion in~\cite{Beneke:2014pta} 
we choose not to expand the prefactor $1/s$ around $1/(4m_b^2)$. Contrary to the 
perturbative contribution, we cannot split the condensate corrections into a resonance 
and continuum part, since both are separately divergent~\cite{Beneke:2014pta}. The total 
corrections to the moments are however well-defined and can be computed numerically using 
the representation of the moments~\eqref{eq:Mn_def} involving contour integration 
or, in principle, analytically by taking derivatives at $q^2=0$. 

\begin{figure}
 \begin{center}
    \includegraphics[height=7.8cm]{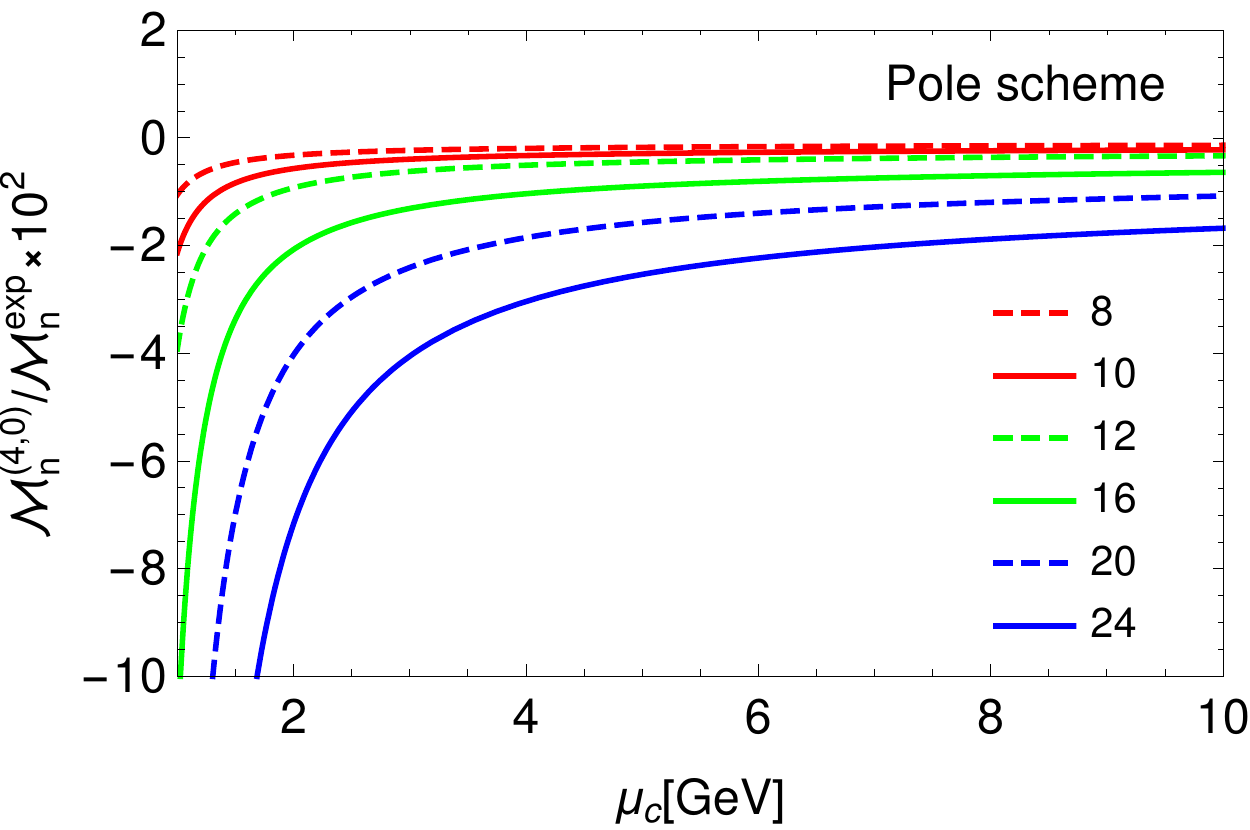}\\[0.8cm]
    \includegraphics[height=7.8cm]{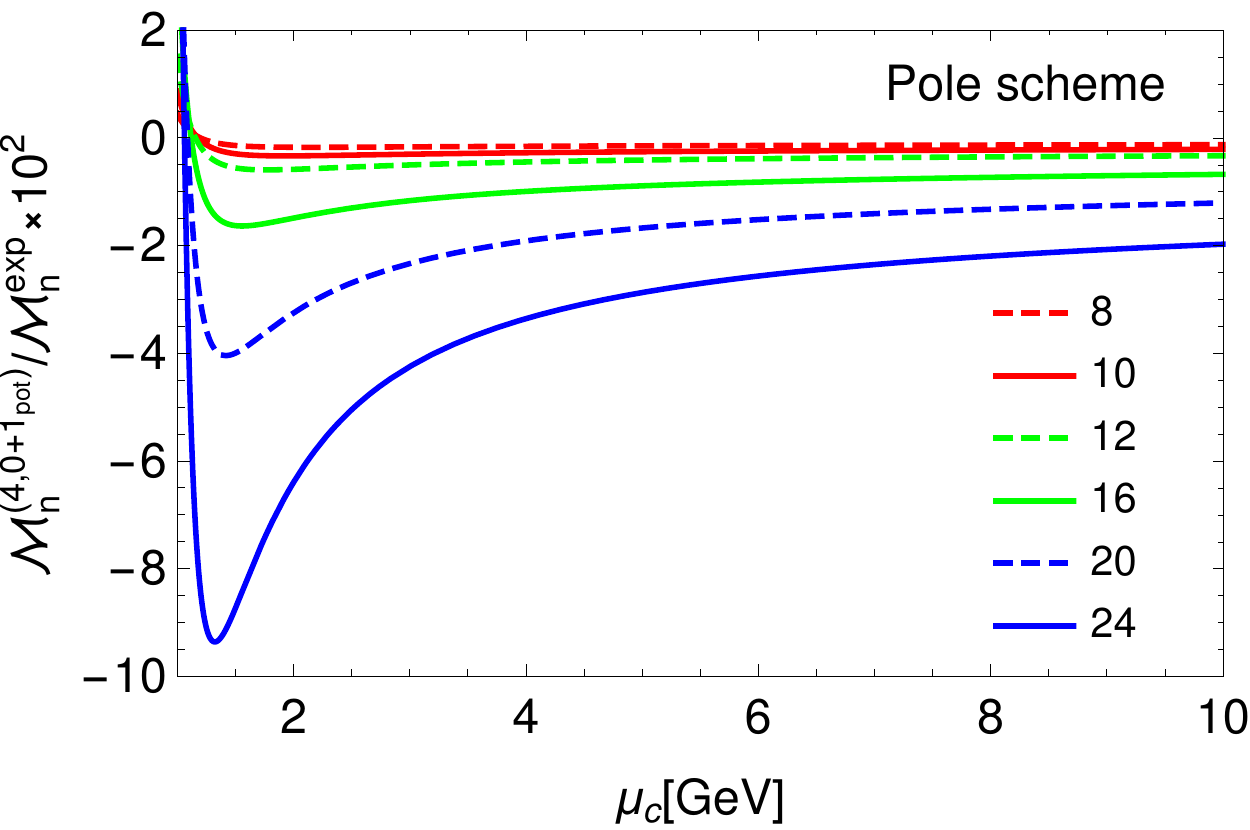}
  \caption{Dimension-four corrections to various non-relativistic moments relative 
  to the experimental moments from~\cite{Beneke:2014pta}. The upper panel shows the 
  leading order contribution and the lower panel the leading order contribution plus 
  the potential corrections. The relative corrections have been rescaled by a factor 
  of 100.}
  \label{fig:Moments_dim4}
 \end{center}
\end{figure}

The scale dependence of the dimension-four contribution are shown in Figure~\ref{fig:Moments_dim4}. 
Results are given in the pole mass scheme using the same inputs as given above. 
We refrain from using the PS or other threshold mass schemes, because the perturbative 
expansion in these schemes becomes unstable in large regions of the scale $\mu_c$. 
This can be traced back to the appearance of large powers of $\lambda$ in the expression 
for the Green function~\eqref{eq:deltaG2G}, which are expanded in the PS mass scheme as 
\begin{equation}
 \lambda^k = \left(\frac{m_b^\text{PS}\alpha_sC_F}{2\sqrt{-m_b^\text{PS} E^\text{PS}}}\right)^k \left(1+\frac{k\,\delta m_b^{\text{PS},\,(1)}}{E^\text{PS}}+\dots\right),
\end{equation}
where $E^\text{PS} = \sqrt{s}-2m_b^\text{PS}-2\delta m_b^{\text{PS},\,(0)}$ and 
$\delta m_b^{\text{PS},\,(i)}$ is the N$^i$LO contribution to the PS-pole mass 
relation.\footnote{However, taken at face value, the dimension-four contributions 
in the PS scheme are smaller than in the pole scheme.} 
This is reminiscent of the destabilization of the NLO correction to the gluon 
condensate contribution~\cite{Broadhurst:1994qj} to the relativistic moments 
in the \MS{} scheme~\cite{Chetyrkin:2010ic}. 

The top panel of Figure~\ref{fig:Moments_dim4} shows the leading order result. 
Although the contribution is proportional to $R_4\propto\alpha_s^{-6}(\mu_c)$, 
its absolute value decreases for larger scales $\mu_c$. Given that the condensate 
corrections to the $\Upsilon(1S)$ mass and leptonic decay rate become unstable 
for scales larger than about 2\,GeV and 1.6\,GeV, respectively, this behaviour 
must be caused by very pronounced cancellations between the contribution from 
the $\Upsilon(1S)$ resonance and the remaining resonances and the continuum (rest), 
which was pointed out in~\cite{Beneke:2014pta}. For the tenth, sixteenth and 
twenty-fourth moment, this cancellation is effective at the level of one part 
in 139, 52 and 20 at the scale $\mu_c=2$\,GeV and at one part in 1530, 659 and 
297 for $\mu_c=10$\,GeV and the growth of the degree of the cancellation for 
higher scales dominates over the growth of the factor $\alpha_s^{-6}(\mu_c)$. 
While this qualitative behaviour is expected due to the reduced infrared 
sensitivity of the moments compared to the properties of the $\Upsilon(1S)$ 
resonance, the extent of the cancellations and the resulting smallness of the 
corrections is rather surprising, especially for larger values of $n\gtrsim16$ 
where power counting predicts a breakdown of the expansion in powers of 
$(n\LQCD/m_b)$. 

The results including the potential NLO corrections are shown in the lower 
panel of Figure~\ref{fig:Moments_dim4}. Above 3\,GeV the corrections do not 
exceed the size of about $\pm$20\% for the considered moments. This is due 
to even more pronounced cancellations within the potential corrections which 
are effective up to about one part in $10^4\,(n=10)$, $5\cdot10^4\,(n=16)$ 
and $2\cdot10^3\,(n=24)$ at $\mu_c=m_b$. The corrections mainly have the 
effect of stabilizing the scale dependence at lower scales $\mu_c\lesssim2$\,GeV, 
such that we find good behaviour of the dimension-four contribution at partial 
NLO over the considered range of scales between 1 and 10\,GeV. 

We can try to assess the convergence of the condensate expansion based only on 
the dimension-four results. Compared to the perturbative result, they are of the 
relative order $1/n\times(n\LQCD/m_b)^4$, where the extra factor of $1/n$ accounts 
for the $v^2$-suppression from the two insertions of the dipole operator. 
Thus, we expect a breakdown of the condensate expansion in $n\LQCD/m_b$ when the 
dimension-four contribution is of the relative size $1/n$. From the lower plot 
in Figure~\ref{fig:Moments_dim4} we deduce that this point is reached in the ballpark 
of $n\approx20$, where the condensate contribution is of the size of -4\% of the 
experimental moment at its peak, which is compatible with the expectation from 
the power counting argument. 

\begin{figure}
 \begin{center}
    \includegraphics[height=7cm]{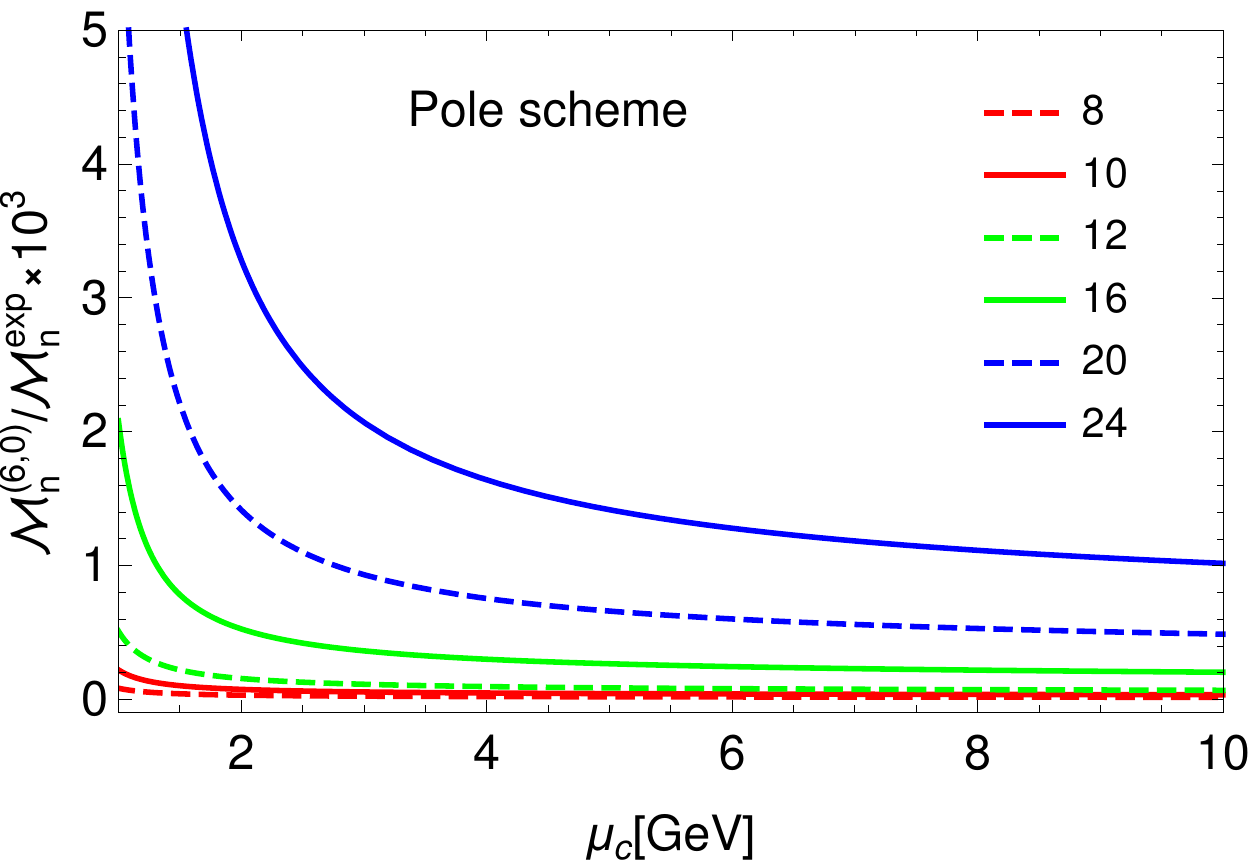}\\[0.8cm]
    \includegraphics[height=7cm]{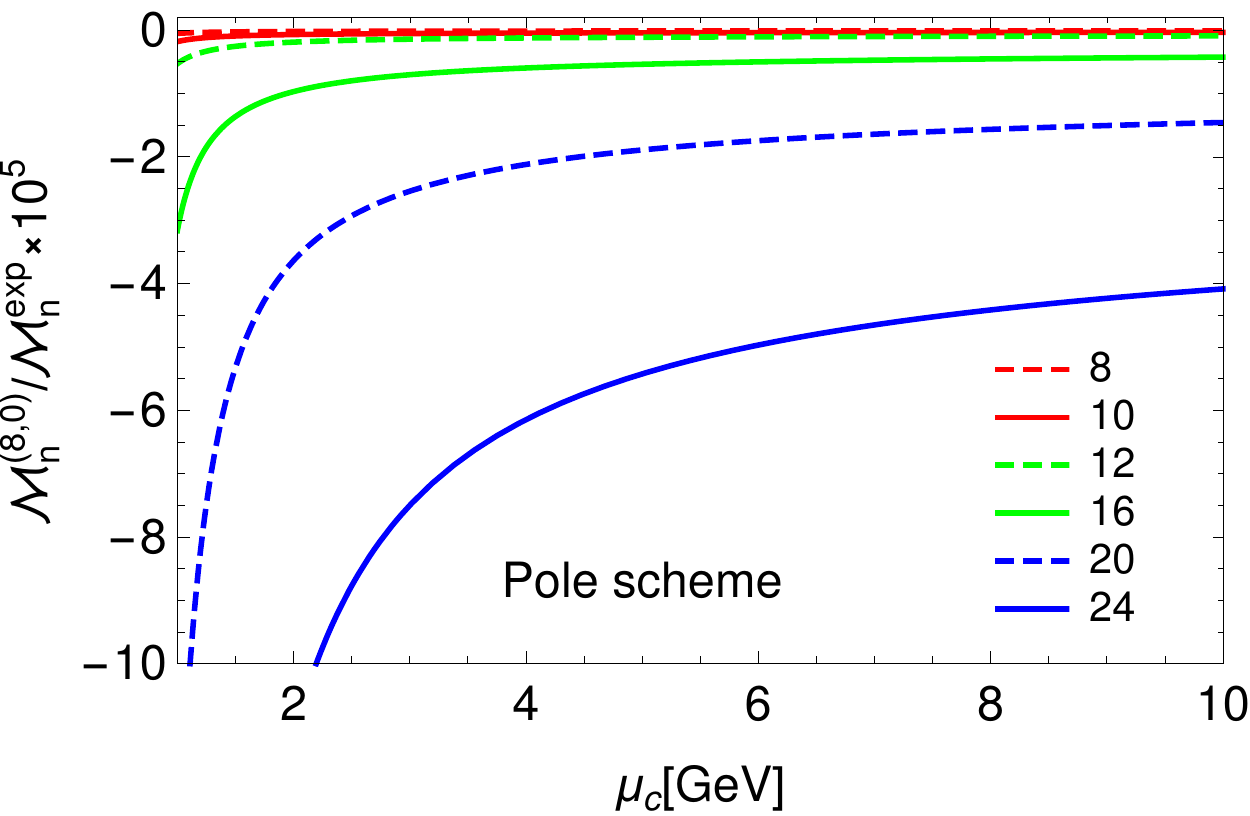}
  \caption{Relative corrections to the non-relativistic moments from the 
  condensate contributions of dimension six (upper panel) and eight (lower panel).
  The relative corrections have been rescaled by factors of $10^3$ and $10^5$, 
  respectively.}
  \label{fig:Moments_dim6_8}
 \end{center}
\end{figure}

In Figure~\ref{fig:Moments_dim6_8}, the relative contributions of dimension six 
(upper panel) and eight (lower panel) are shown. Both are significantly smaller 
than our expectation based on the putative breakdown of the expansion 
around $n\approx20$, which would imply that the dimension six and eight corrections 
are both of the order $1/n\approx0.05$. This smallness is the result of cancellations 
between the contribution from the $\Upsilon(1S)$ resonance and the rest that are 
even stronger than at dimension four. Explicitly, they are at the level of about 
one part in $3\cdot10^5\,(n=10)$, $5\cdot10^4\,(n=16)$ and $10^4\,(n=24)$ at dimension 
six and about one part in $10^8\,(n=10)$, $10^7\,(n=16)$ and $2\cdot10^6\,(n=24)$ 
at dimension eight. We believe that the reason for this behaviour is the off-shellness 
of the moments which are defined as derivatives of the vacuum polarization function 
at $q^2=0$, far away from the physical cut at $s\geq M_{\Upsilon(1S)}^2$. 
This off-shellness effectively acts as an IR cutoff and suppresses higher-dimensional 
corrections, which probe the IR regime. On the other hand, the properties of the Upsilon 
resonances, that we discussed above, are on-shell quantities and the higher-dimensional 
condensate contributions do not appear suppressed with respect to our expectations from 
power counting. 

From the point of view of the convergence of the condensate expansion, it appears 
that the moments can be described reliably up to values of $n$ much larger than 20. 
However, as pointed out in~\cite{Beneke:2014pta}, the validity of quark-hadron 
duality must be questioned when the moment is completely saturated by lowest state. 
This is the case for the higher values considered here, where the relative contribution 
of the $\Upsilon(1S)$ to the experimental moments amounts to 95\% for $n=20$ and 97\% 
for $n=24$~\cite{Beneke:2014pta}. By the term 'violation of quark-hadron duality' 
we refer to contributions which have a trivial Taylor expansion and are, therefore, 
not captured by the condensate expansion, like e.g. exponential terms of the form 
$\text{exp}(-m_b/(n\LQCD))$. Behaviour that is consistent with the presence of 
such contributions has been observed in the 't Hooft model~\cite{Bigi:1998kc}.\footnote{
Ref. \cite{Bigi:1998kc} considers observables in the Minkowski domain, where the 
exponential terms must be analytically continued and manifest as oscillations.} 
However, the size of these contributions in four-dimensional QCD is difficult to 
quantify and we do not attempt this here. We note, however, that the exponential 
terms originate from coherent soft fluctuations~\cite{Shifman:2000jv}, e.g. from 
contributions where the off-shellness is distributed among many soft lines carrying 
momenta of the order $\LQCD$, which pushes the bottom pair close to its mass shell. 
It is conceivable that such an effect does not experience a similar suppression 
from the effective IR cutoff as the higher-dimensional condensate contributions. 

In the range $n\LQCD\sim m_b$ the exponential $\text{exp}(-m_b/(n\LQCD))$ is of order 
one and we cannot exclude that duality violation effects are relevant at the high 
accuracy we require for reliable determinations of the bottom-quark mass. 
We conclude that, in practice, the range of moments is limited by our knowledge 
of the validity of quark-hadron duality and not by the convergence of the condensate 
expansion and advise that moments with $n\gtrsim16$ are not used for determinations 
of the bottom-quark mass. On the other hand, for $n\approx10$ duality-violating 
effects are exponentially suppressed and the condensate expansion provides a reliable 
determination of the non-perturbative effects. Our results given in 
Figure~\ref{fig:Moments_dim4} and \ref{fig:Moments_dim6_8} show that the condensate 
contributions in this region are in the subpercent range and can safely be neglected 
compared to the perturbative uncertainties.

%%%%%%%%%%%%%%%%%%%%%%%%%%%%%%%%%%%%%%%%%%%%%%%%%%%%%%%%%%%%%%%%%%%%%%%%%%%%%%%%%%%%%%%%%%
%%%%%%%%%%%%%%%%%%%%%%%%%%%%%%%%%     Conclusions      %%%%%%%%%%%%%%%%%%%%%%%%%%%%%%%%%%%
%%%%%%%%%%%%%%%%%%%%%%%%%%%%%%%%%%%%%%%%%%%%%%%%%%%%%%%%%%%%%%%%%%%%%%%%%%%%%%%%%%%%%%%%%%

\section{Conclusions\label{sec:conclusions}}

We have determined the leading order condensate corrections to the $\Upsilon(NS)$ 
masses, leptonic decay rates and sum rules up to and including dimension eight. 
In addition the potential NLO corrections to the dimension-four contribution 
have been computed, which allows us to assess the preferred scale choice in the 
condensate corrections. Our results suggest that the expansion is well behaved 
for the mass of the $\Upsilon(1S)$, but breaks down for the higher states. 
The former observation has been used to determine the bottom-quark mass with 
the results given in~\eqref{eq:mbPS} and \eqref{eq:mbMS}. 

The leading order condensate corrections to $\Gamma(\Upsilon(1S)\to l^+l^-)$ 
have a small window of convergence for $0.8\,\text{GeV}\lesssim\mu_c\lesssim1.6\,\text{GeV}$, 
where they lead to good agreement with the experimental value, but the partial NLO 
corrections to the dimension-four contribution exceed the leading order correction 
and cause us to question the perturbative stability. Thus, a final verdict for the 
leptonic decay rate of the $\Upsilon(1S)$ is only possible once the missing ultrasoft 
correction has been calculated. 

Last but not least, we have considered the non-relativistic moments~\eqref{eq:Mn_def}. 
We find extremely good convergence of the higher-dimensional condensate contributions 
which clearly shows that non-perturbative contributions to moments with $n\approx10$ 
are negligible. On the other hand, we cannot unambiguously exclude the possibility of 
relevant violations of of quark-hadron duality for $n\gtrsim16$ despite the surprising 
smallness of the dimension six and eight corrections. Thus, the non-relativistic 
moments with $n\approx10$ remain the theoretically cleanest approach for determinations 
of the bottom-quark mass from the $\Upsilon$ system.

\subsection*{Acknowledgements}

I am grateful to M.~Beneke, A.~Maier and M.~Stahlhofen for helpful discussions, 
to A.~Maier for comments on the manuscript and to 
V.~Mateu and P.~Ortega for communication regarding~\cite{Mateu:2017hlz}. 
I wish to thank the Erwin Schr\"odinger International Institute for Mathematics and Physics (ESI) 
in Vienna for hospitality during the programme \emph{Challenges and Concepts for Field Theory 
and Applications in the Era of LHC Run-2} where part of this work was done.

\appendix

%%%%%%%%%%%%%%%%%%%%%%%%%%%%%%%%%%%%%%%%%%%%%%%%%%%%%%%%%%%%%%%%%%%%%%%%%%%%%%%%%%%%%%%%%%
%%%%%%%%%%%%%%%%%%%%%%     energy levels and wave functions      %%%%%%%%%%%%%%%%%%%%%%%%%
%%%%%%%%%%%%%%%%%%%%%%%%%%%%%%%%%%%%%%%%%%%%%%%%%%%%%%%%%%%%%%%%%%%%%%%%%%%%%%%%%%%%%%%%%%

\section{Condensate corrections to the energy levels and wave functions\label{sec:eANDf}}

We give the results for the condensate corrections to the energy levels and the 
wave functions at the origin of the S-wave bottomonium states. The contributions 
are parametrized as 
\begin{eqnarray}
 E_N & = & E_N^{(0)} \left(1 + \sum\limits_{i=1}^\infty e_N^{(i)} + \sum\limits_{k=4,6,8,\dots}\sum\limits_{l=0}^\infty e_N^{(k,l)}\right),\\
 |\psi_N(0)|^2 & = & |\psi_N^{(0)}(0)|^2 \left(1 + \sum\limits_{i=1}^\infty f_N^{(i)} + \sum\limits_{k=4,6,8,\dots}\sum\limits_{l=0}^\infty f_N^{(k,l)}\right),
\end{eqnarray}
where the leading order expressions are given by 
\begin{equation}
 E_n^{(0)} = -\frac{m_b\alpha_s^2C_F^2}{4n^2},\hspace{1.5cm}|\psi_n^{(0)}(0)|^2 = \frac{1}{\pi}\left(\frac{m_b\alpha_sC_F}{2n}\right)^3,
\end{equation}
the perturbative corrections of relative order $\alpha_s^i$ are $e_N^{(i)}$ and $f_N^{(i)}$, 
and $e_N^{(k,l)}, f_N^{(k,l)}$ are the condensate corrections of relative order 
$(\LQCD/(m_b\alpha_s^2))^k \alpha_s^{l+2}$ to the $N$th energy level and wave function, 
respectively. At dimension four, we obtain 
\begin{eqnarray}
 e_N^{(4,0)} & = &  R_4 \, \frac{32 N^6 \left(25515 N^6-109935 N^4+101216 N^2-26624\right)}{6561 N^4-25920 N^2+16384},\\
 f_N^{(4,0)} & = & -R_4 \, \frac{32N^6}{9\left(6561 N^4-25920 N^2+16384\right)^2}\nonumber\\
                              &   & \times\Big(4519905705 N^{10}-36791660430 N^8+101725313184 N^6\nonumber\\
                              &   & -112065638400 N^4+50981371904 N^2-7583301632\Big),
\end{eqnarray}
in agreement with the results from~\cite{Voloshin:1979uv,Leutwyler:1980tn}. 
For the dimension-six corrections to the energy levels and wave functions, we find 
\begin{eqnarray}
 e_N^{(6,0)} & = & R_6\,\frac{4096 N^{10}}{81 \left(9 N^2-64\right) \left(6561 N^4-25920 N^2+16384\right)^3}\nonumber\\
             &   & \times\big[65241222927111 N^{16}-1327743092409993 N^{14}+10789755579716526 N^{12}\nonumber\\
             &   & -46158344158975776 N^{10}+114216987240880128 N^8\nonumber\\
             &   & -168309372752363520 N^6+145600287615221760 N^4\nonumber\\
             &   & -68153404341354496 N^2+13295844358881280\big],\\
 f_N^{(6,0)} & = & R_6\,\frac{4096 N^{10}}{81 (64 - 9 N^2)^2 (1024 - 81 N^2) (6561 N^4 - 25920 N^2 + 16384)^4}\nonumber\\
             &   & \times\big[1560233733912305862795 N^{24}-68302865242974003997572 N^{22}\nonumber\\
             &   & +1257835587897041879681466 N^{20}-12922847845013954087408448 N^{18}\nonumber\\
             &   & +82659284132080163141376000 N^{16}-347414281805040198547931136 N^{14}\nonumber\\
             &   & +985563190675064665304727552 N^{12}-1906052104684436293825855488 N^{10}\nonumber\\
             &   & +2504628423489707401549971456 N^8-2195117868501221112538988544 N^6\nonumber\\
             &   & +1227049495579909701471567872 N^4-395519535823226068598259712 N^2\nonumber\\
             &   & +55919902706900903797981184\big].
             \label{eq:e6f6}
\end{eqnarray}
The correction to the energy levels is identical to the result 
of~\cite{Pineda:1996uk}. Our result for the wave function correction 
however differs from the one in~\cite{Pineda:1996uk} in the 
coefficients in the square bracket that multiply powers of $N$, 
while the constant term is in agreement. Numerically the difference is 
tiny, dropping from 3 permille for $N=1$ to 1.8 permille for $N=10$. 
Our dimension-eight results read 
\begin{eqnarray}
\label{eq:e8}
 e_N^{(8,0)} & = & R_8\,\frac{131072 N^{14}}{6561 (64 - 9 N^2)^3 (1024 - 81 N^2) (16384 - 25920 N^2 + 6561 N^4)^5}\\
             &   & \hspace{-1cm} \times\big[
                    513297061199674600970728035 N^{30}-28809695301605440114072286106 N^{28}\nonumber\\
             &   & \hspace{-1cm} +714217935339861807140929892403 N^{26}-10391224399757404953006517310268 N^{24}\nonumber\\
             &   & \hspace{-1cm} +99331264481676577483010819164416 N^{22}-661340018691822569991363819749376 N^{20}\nonumber\\
             &   & \hspace{-1cm} +3169778592264419379462212875714560 N^{18}-11139087489514263003228295894401024 N^{16}\nonumber\\
             &   & \hspace{-1cm} +28931645127160026102581319759298560 N^{14}-55513215835612849636269917466525696 N^{12}\nonumber\\
             &   & \hspace{-1cm} +78003482654565522219373939233128448 N^{10}-78799349232622194122863330417704960 N^8\nonumber\\
             &   & \hspace{-1cm} +55418913930949175463048951864754176 N^6-25648900900141403595678833574936576 N^4\nonumber\\
             &   & \hspace{-1cm} +6999998028019916122574328222973952 N^2-851560509652109050320418236268544 \big],\nonumber
\end{eqnarray}
\begin{eqnarray}
 f_N^{(8,0)} & = & R_8\,\frac{-131072 N^{14}\,(1600 - 81 N^2)^{-2}(256 - 9 N^2)^{-1}}{59049  (64 - 9 N^2)^4 (1024 - 81 N^2)^3 (16384 - 25920 N^2 + 6561 N^4)^6}\nonumber\\
             &   & \times\big[739783191218801346196996467082948404493005 N^{46}\nonumber\\
             &   & -118424806386048034335763849263957780781041084 N^{44}\nonumber\\
             &   & +8614327589477425734307468706116579381193741895 N^{42}\nonumber\\
             &   & -378783092051612274031903244647052432202630343536 N^{40}\nonumber\\
             &   & +11297942118461809329963818184791182730335090044672 N^{38}\nonumber\\
             &   & -243143128750125652415373537739611048866722611757056 N^{36}\nonumber\\
             &   & +3920817963564181189981156819150310468833247662964736 N^{34}\nonumber\\
             &   & -48564520997046676673717761034578880399807666580357120 N^{32}\nonumber\\
             &   & +469935568662507411174423052932263135345073787276099584 N^{30}\nonumber\\
             &   & -3594124289377594431764429130697307804693677753786957824 N^{28}\nonumber\\
             &   & +21896983904235671174732114490714870006523720168378466304 N^{26}\nonumber\\
             &   & -106786420612452063591697653002182767135747603942793543680 N^{24}\nonumber\\
             &   & +417824025967497761694434177318848448549510813159365017600 N^{22}\nonumber\\
             &   & -1311624595310101793819336380085499993464121204574984863744 N^{20}\nonumber\\
             &   & +3295683076732204940723945262255632915195793616243217399808 N^{18}\nonumber\\
             &   & -6596089628312909075581437831841936847452213443718287458304 N^{16}\nonumber\\
             &   & +10433707912422547412099437180603984258749322917926072745984 N^{14}\nonumber\\
             &   & -12894168869874165524006895663064691969753810987628940492800 N^{12}\nonumber\\
             &   & +12244261212399551038854557341492532616026313680323426123776 N^{10}\nonumber\\
             &   & -8720306429636174559372052451043628307310497364671742345216 N^8\nonumber\\
             &   & +4489452005081410411327292504232625921839722110283767873536 N^6\nonumber\\
             &   & -1572764930888520531444116515121465620114569888243055067136 N^4\nonumber\\
             &   & +334662695930263285190103020209714815173408421809068441600 N^2\nonumber\\
             &   & -32577921122492492137783209690205922464480693174927360000 \big].
\label{eq:f8}
\end{eqnarray}

The NLO corrections to the dimension-four condensate contributions have the form 
\begin{eqnarray}
\label{eq:eN41}
 e_N^{(4,1)} & = & R_4\,\frac{\alpha_s}{4\pi}\left[\frac{C_A/2-C_F}{-C_F}\widetilde{e}_N^\text{ DVD}+\widetilde{e}_N^\text{ DDV}+\widetilde{e}_N^\text{ us}\right],\\
 f_N^{(4,1)} & = & R_4\,\frac{\alpha_s}{4\pi}\left[\frac{C_A/2-C_F}{-C_F}\widetilde{f}_N^\text{ DVD}+\widetilde{f}_N^\text{ DDV}+\widetilde{f}_N^\text{ us}\right],
\label{eq:fN41}
\end{eqnarray}
where the potential terms are
\begin{eqnarray}
 \widetilde{e},_N^\text{ DVD} & = & \frac{128 N^6}{9 \left(6561 N^4-25920 N^2+16384\right)^2}\Bigg\{2 \left[a_1+2 \beta_0 \left(S_1+L_N\right)\right]\Big(167403915 N^{10} \nonumber\\
 & & -1486558575 N^8+4690934208 N^6-6303780864 N^4+3483631616 N^2 \nonumber\\
 & & -536870912\Big)+\beta_0(9N-8)\Big(94419351 N^9+46727442 N^8-800382951 N^7 \nonumber\\
 & & -381105162 N^6+2367378684 N^5+1061906784 N^4-2883647232 N^3 \nonumber\\
 & & -1162401792 N^2+1320550400 N+405274624\Big)\Bigg\},
\end{eqnarray}
\begin{eqnarray}
 \widetilde{e},_N^\text{ DDV} & = & \left[a_1+2 \beta_0 \left(S_1+L_N\right)\right]\frac{-32 N^6 }{9 \left(6561 N^4-25920 N^2+16384\right)^2}\Big(5859137025 N^{10} \nonumber\\
 & & -48288205485 N^8+136847786688 N^6-159273676800 N^4 \nonumber\\
 & & +81059381248 N^2-15166603264\Big)+\nonumber\\
 & & \beta_0\frac{32 N^6 }{3 \left(6561 N^4-25920 N^2+16384\right)}\Big(451332 N^6-161595 N^5-1797839 N^4 \nonumber\\
 & & +646725 N^3+1160272 N^2-425280 N+7168\Big),
\end{eqnarray}
\begin{eqnarray}
 \widetilde{f},_N^\text{ DVD} & = & \left(a_1+2 \beta_0 L_N\right)\frac{256 N^6}{81 \left(6561 N^4-25920 N^2+16384\right)^3}\Big(-29655101330505 N^{14} \nonumber\\
 & & +370819521046350 N^{12}-1836925493383872 N^{10}+4598115283537920 N^8 \nonumber\\
 & & -6194386829574144 N^6+4467901285269504 N^4-1563638678683648 N^2 \nonumber\\
 & & +189115999977472\Big)+\beta_0\frac{128 N^5}{81 \left(6561 N^4-25920 N^2+16384\right)^2}\Bigg\{ \nonumber\\
 & & \frac{1}{9 \left(6561 N^4-25920 N^2+16384\right)^2}\Big(-8888937845922281277 N^{19} \nonumber\\
 & & +2334805437953319660 N^{18}+146679930871209367335 N^{17} \nonumber\\
 & & -38038386654486402750 N^{16}-1019696917323083770998 N^{15} \nonumber\\
 & & +260020266583548188160 N^{14}+3893749911649043582976 N^{13} \nonumber\\
 & & -970389752080885235712 N^{12}-8954749600639565709312 N^{11} \nonumber\\
 & & +2161122845010608259072 N^{10}+12847315095944310030336 N^9 \nonumber\\
 & & -2955869597186629042176 N^8-11551005743863723720704 N^7 \nonumber\\
 & & +2436883078824792686592 N^6+6290577008666263683072 N^5 \nonumber\\
 & & -1065043454471757103104 N^4-1829313053453482721280 N^3 \nonumber\\
 & & +131764347572768997376 N^2+182269121168985292800 N \nonumber\\
 & & +36893488147419103232\Big)+\frac{36 N  S_1}{6561 N^4-25920 N^2+16384} \nonumber\\
 & & \times\Big(-3295011258945 N^{14}+41202169005150 N^{12}-204102832598208 N^{10} \nonumber\\
 & & +510901698170880 N^8-688295779762176 N^6+498462848188416 N^4 \nonumber\\
 & & -178391466639360 N^2+21990232555520\Big)-\frac{33554432 N}{6561 N^4-25920 N^2+16384} \nonumber\\
 & & \times\Bigg[\left(-32805 N^6+2177280 N^4-4993024 N^2+1048576\right) S_1\left(\frac{N}{8}\right) \nonumber\\
 & & +\left(-111537 N^6+124416 N^4-5386240 N^2+7340032\right) S_1\left(\frac{9N}{8}\right)\Bigg] \nonumber\\
 & & -36 N^2\left(S_2-\frac{\pi ^2}{6}\right) \Big(-167403915 N^{10}+1486558575 N^8-4690934208 N^6 \nonumber\\
 & & +6303780864 N^4-3483631616 N^2+536870912\Big) \Bigg\},
\end{eqnarray}
\begin{eqnarray}
 \widetilde{f},_N^\text{ DDV} & = & \left(a_1+2 \beta_0 L_N\right)\frac{-64 N^6}{81 \left(6561 N^4-25920 N^2+16384\right)^3}\Big(-385516317296565 N^{14} \nonumber\\
 & & +4654961778867030 N^{12}-21965572921731408 N^{10}+51360436390947840 N^8 \nonumber\\
 & & -63132424989769728 N^6+41064256744980480 N^4-13147993330941952 N^2 \nonumber\\
 & & +1582746988183552\Big)-\beta_0\frac{32 N^5}{81 \left(6561 N^4-25920 N^2+16384\right)^2}\Big\{ \nonumber\\
 & & \frac{1}{9 \left(6561 N^4-25920 N^2+16384\right)^2}\Big(61405383018172307058 N^{19} \nonumber\\
 & & -22180651660556536770 N^{18}-965837565458451500139 N^{17} \nonumber\\
 & & +351080411169493108575 N^{16}+6289521884171512783128 N^{15} \nonumber\\
 & & -2307675798943664843520 N^{14}-21902675740289714221056 N^{13} \nonumber\\
 & & +8153996687233488691200 N^{12}+44045654765398681780224 N^{11} \nonumber\\
 & & -16799371281675894915072 N^{10}-51685121285332515422208 N^9 \nonumber\\
 & & +20621242559717072437248 N^8+33974735719419655225344 N^7 \nonumber\\
 & & -14944437608241652826112 N^6-10679923278040006656000 N^5 \nonumber\\
 & & +6036677324559894970368 N^4+609631074176867500032 N^3 \nonumber\\
 & & -1054643390588414590976 N^2+220638101144321654784 N \nonumber\\
 & & -18446744073709551616\Big)+\frac{4 N S_1}{6561 N^4-25920 N^2+16384} \nonumber\\
 & & \times\Big(-652412229271110 N^{14}+7881876138460500 N^{12}-37221596700680304 N^{10} \nonumber\\
 & & +87133432412759040 N^8-107286357009825792 N^6+69985956437950464 N^4 \nonumber\\
 & & -22538834096947200 N^2+2696552267120640\Big)+\frac{16777216 N }{6561 N^4-25920 N^2+16384} \nonumber\\
 & & \Big(\left(203391 N^6-13405824 N^4+24866816 N^2+1048576\right) S_1\left(\frac{N}{8}\right) \nonumber\\
 & & +\left(-465831 N^6+30824064 N^4-64811008 N^2+7340032\right) S_1\left(\frac{9N}{8}\right)\Big) \nonumber\\
 & & -18 N^2\left(S_2-\frac{\pi ^2}{6}\right) \Big(-5859137025 N^{10}+48288205485 N^8-136847786688 N^6 \nonumber\\
 & & +159273676800 N^4-81050992640 N^2+14629732352\Big) \Bigg\},
\end{eqnarray}
where $L_N=\ln(N\mu/(m\alpha_sC_F))$, $S_1(x) = \sum_{k=1}^x k^{-1}$ 
is the analytic continuation of the harmonic number to non-integer values 
and $S_i = \sum_{k=1}^N k^{-i}$ without an explicit argument is the 
$N$th harmonic number of rank $i$. The ultrasoft corrections $\widetilde{e}_N^\text{ us}$ 
and $\widetilde{f}_N^\text{ us}$ are currently unknown.

\end{document}